\documentclass[12pt,twoside]{article} 
\usepackage[top=30mm, bottom=30mm, left=25mm, right=25mm]{geometry}

\usepackage[utf8]{inputenc}
\usepackage{amsmath,amsthm,amssymb,amsfonts}
\usepackage[english]{babel}
\usepackage{enumitem}
\usepackage[T1]{fontenc}
\usepackage{graphicx}
\usepackage{dsfont}
\usepackage{float}
\usepackage{caption}
\usepackage{color}
\usepackage[colorlinks,linkcolor=blue]{hyperref}

\newcommand{\pjk}{\psi_{j,k} } %
\newcommand{\BE}{\begin{equation}}
\newcommand{\la}{\lambda}
\newcommand{\EE}{\end{equation}}
\newcommand{\TPA}{T_{\alpha}^p (x_0)}
\newcommand{\HFP}{h_{f}^p (x_0)}

\newcommand{\RR}{{\mathbb R}}

\newcommand{\ZZ}{{\mathbb Z}}
\newcommand{\NN}{{\mathbb N}}

\newcommand{\ep}{\varepsilon}

\newcommand{\Hmin}{H^{\text{min}}_f}
\newcommand{\ome}{\omega}
\newcommand{\dws}{{\mathcal D}^{ws}_f}
\newcommand{\djk}{d_{j,k}}
\newcommand{\cjk}{c_{j,k}}
\newcommand{\deq}{:=} 
 \newcommand{\Proba}{\mathbb{P}} 
\newcommand{\paren}[1]{{\left( #1 \right)}} %
\newcommand{\abs}[1]{{\left\lvert#1\right\rvert}} %
\newcommand{\brho}{\bsb{\rho}}
\newcommand{\al}{\alpha} %
\newcommand{\bhmin}{{H^{\min}_X}}
\newcommand{\bhmax}{{H^{\max}_X}}

\newtheorem{defi}{Definition}[section]
\newtheorem{rem}{Remark}[section]
\newtheorem{theorem}{Theorem}[section]
\newtheorem{coro}{Corollary}[section]
\newtheorem{lem}{Lemma}[section]
\newtheorem{prop}{Proposition}[section]

\newcommand{\BD}{\begin{defi}}
\newcommand{\ED}{\end{defi}}
\newcommand{\BR}{\begin{rem}}
\newcommand{\ER}{\end{rem}}
\newcommand{\BC}{\begin{coro}}
\newcommand{\EC}{\end{coro}}
\newcommand{\BL}{\begin{lem}}
\newcommand{\EL}{\end{lem}}
\newcommand{\BP}{\begin{prop}}
\newcommand{\EP}{\end{prop}}
\newcommand{\BT}{\begin{theorem}}
\newcommand{\ET}{\end{theorem}}
\newcommand{\bsb}[1]{\boldsymbol{#1}} 

\newsavebox{\fmbox}
\newenvironment{fmpage}[1]
 {\begin{lrbox}{\fmbox}\begin{minipage}{#1}}
 {\end{minipage}\end{lrbox}\fbox{\usebox{\fmbox}}}

\title{Multifractal analysis based on  weak scaling exponents:  Applications to MEG recordings in neuroscience}

\author{Merlin Dumeur\thanks{CEA (NeuroSpin), Inria MIND, Gif-sur-Yvette, France \\  \texttt{merlin.dumeur@protonmail.com}, \texttt{philippe.ciuciu@cea.fr}}, Guillaume Saës\thanks{Univ Paris Est Creteil, Univ Gustave Eiffel CNRS, LAMA UMR8050, F-94010 Creteil, France \\  \texttt{guillaume.saes@u-pec.fr}, \texttt{jaffard@u-pec.fr}},
    Patrice Abry\thanks{CNRS, ENS de Lyon, LPENSL, UMR5672, 69342, Lyon cedex 07, France \\ \texttt{patrice.abry@ens-lyon.fr}}, \\ 
    Philippe Ciuciu$^{\ast}$, Herwig Wendt\thanks{CNRS, IRIT, Université de Toulouse, Toulouse, France\\  \texttt{herwig.wendt@irit.fr}},
    Stéphane Jaffard$^{\dagger}$ 
}

\date{}

\begin{document}

\maketitle

{\textbf{Abstract }}: 
A novel multifractal analysis  based on the weak scaling exponents is developed and its mathematical properties are studied.
A key advantage, compared to earlier formulations based on H\"older or $p$-exponents, consists of the fact that it does not rely on the assumption of any a priori global regularity. 
To illustrate its potential in real world applications, we show that this method allows to study the regularity of MEG signals, recording electromagnetic brain activity, which was not possible using the formerly introduced methods based on H\"older or $p$-exponents, without preprocessing. 

\clearpage
\newpage
\tableofcontents
\clearpage
\newpage

{\em This article is dedicated to Akram Aldroubi: a true friend,  a great scientist,  and a fantastic salsa dancer!}

\section{Introduction}

The purpose of this article is to develop the mathematical understanding of a variant of multifractal analysis, which does not require a priori regularity assumptions on the data to be analyzed, in contrast to all other multifractal analysis methods introduced in the past; furthermore, we show it at work on MEG signals, which record electromagnetic brain activity from SQUID sensors located around the patient's head. 
The reason for testing this new framework on such data is that MEG signals often do not meet the a priori regularity assumptions required by other methods.  
We start by recalling the purpose and aims of multifractal analysis  from a signal processing viewpoint. 

\subsection{Multifractal analysis} 

Multifractal analysis supplies methods which associate to everywhere irregular signals classification parameters based on scaling invariance properties. It can be traced backed to the seminal work of N. Kolmogorov in the 1940s  \cite{k41}  where  the {\em Kolmogorov scaling function} $\zeta_f (q)$ of a function $f$ was introduced as 
\BE\label{autosimkolmo}
\forall q >0, \qquad \int |  f(x+h ) -f(x)|^q dx \sim |h|^{\zeta_f (q)}
\EE
in the limit of small scales $h \rightarrow 0$ (a more precise, but less eloquent definition is
supplied by \eqref{foncech}).  A first success of this  tool is that it  allowed to discard the possibility of modeling the velocity of fully developed turbulence at small scales by fractional Brownian motion (fBm) ; indeed this process has a linear scaling function, which is not the case for turbulence data, see  \cite{Frisch1995} and references therein.  Key steps concerning the understanding of the information supplied by the scaling function were obtained as a consequence of key ideas introduced by  U. Frisch and G. Parisi  in 1985 \cite{ParFri85}: They interpreted the strict concavity of the scaling function as indicating the presence of different values taken  by the pointwise regularity of the function analyzed. Let us be more precise: The pointwise Hölder exponent  of a locally bounded function $f: \RR \rightarrow \RR $  is defined as follows.

\BD
Let $f\in L_{\text{loc}}^{\infty} (\RR)$. Let $x_0\in\RR$ and $\alpha\geq 0$;  $f$ belongs to $C^\alpha (x_0)$ if  there exist a polynomial $P_{f,x_0}$ of degree less than $\alpha$ and $C,r>0$ such that
$$
\forall x \in (x_0-r,x_0+r), \quad |f(x)-P_{f,x_0}(x-x_0)| \leq C |x-x_0|^{\alpha}.
$$
The Hölder exponent of $f$ at $x_0$ is $h_f (x_0)=\sup\{\alpha \ : \ f\in C^{\alpha} (x_0)\}$.

The { multifractal spectrum} of  $f$ is 
\BE \label{defspec}  {\mathcal D}_f (H) = \dim \;  (  \left\{ x : \hspace{3mm} h_f (x) =H\right\}  ) , \EE 
where  $ \dim $ denotes the Hausdorff dimension (and, by convention $\dim (\emptyset) = -\infty $). 
The support of the spectrum is the set of values $H$ for which ${\mathcal D}_f (H)  \neq -\infty$.
\ED
 
The idea underlying the definition of the  multifractal spectrum is that, for large classes of signals, pointwise regularity  varies from point to point in an extremely irregular way, so that its precise determination for each possible location is not a realistic goal, and one should rather focus on estimating more global quantities, such as  the size of the sets of points where a given type of singularity shows up. 
Inspired by the thermodynamics formalism, U. Frisch and G. Parisi proposed a formula for estimating ${\mathcal D}_f (H)$ by means of a Legendre transform  of $\zeta_f $: they defined the { \em Legendre spectrum  } of $f$ as 
\BE \label{formfetp}
{\mathcal L}_f(H) : =  \inf_{q }  \left( 1+Hq - \zeta_f (q)  \right),
\EE
and they developed heuristic arguments backing the idea that, in general, the Legendre spectrum coincides with the multifractal spectrum: when it is the case, the so-called { \em multifractal formalism } is said to hold.
Multifractal analysis (using several possible variants for the definition of the scaling function) has been tried and tested in numerous applications ranging from medical image processing \cite{Gerasimova14, Villain19} to the modeling and prediction of  natural phenomena \cite{Friedrich22, Lashermes08, Robert08} and brain activity in neuroscience~\cite{Ciuciu08,Ciuciu12,la2018self,domingues2019multifractal,dumeur2023multifractality} (see \cite{MandMemor} for a review). 
 
However, two  limitations of the multifractal formalism quickly appeared: 
First it is irrelevant for functions that are not locally bounded, in which case the H\"older exponent is no longer defined.  
This raised the problem of determining if the Kolmogorov scaling function yields some information for other concepts of pointwise regularity; we will come back to this question in Sec.~\ref{sec2}.
A second concern was the numerical instability of the computation of the scaling function when extended to negative values of $q$; this is critical because, restricting the use of \eqref{formfetp} to $q>0$ yields at best the increasing hull of the multifractal spectrum.  
This problem already appeared for models as simple as the Brownian motion. 
Since its pointwise H\"older exponent takes the  constant value $h_B (x) = 1/2$, its multifractal spectrum  is supported by the unique value $H = 1/2$, whereas \eqref{formfetp} yields  a wrong decreasing part for $ {\mathcal D}_B$: Its right hand side takes the value  $3/2 -H$ for $H \in [1/2, 3/2]$, see \cite{JAFFARD:2006:A,Farf4}. 
These limitations  motivated several  advances: 
\begin{itemize}
    \item This setting was soon extended to the analysis of probability measures:  In that case, the pointwise regularity exponent $h_\mu$ of a measure $\mu$ is (informally) defined by 
    \BE \label{mesreg}
    \mu ([x-r, x+r]) \sim r^{h_\mu (x)} \qquad \mbox{ when} \quad h \rightarrow 0 ;
    \EE
    in  1992,  G. Brown, G.  Michon and J. Peyri\`ere proved that the corresponding formalism  (obtained by adapting \eqref{formfetp} to a relevant scaling function, such as \eqref{foncdech} below) yields an upper bound for the multifractal spectrum when the infimum is taken on all (positive and negative) values of $q$, see \cite{Bmp92}. 

     \item As regards  functions, in order to eliminate the numerical instabilities met for $q <0$, A. Arneodo and collaborators introduced an alternative way to compute the  scaling function: In  1991, they proposed to  replace increments in~\eqref{autosimkolmo} by a continuous wavelet transform, and the integral by a discrete sum computed over the local maxima of this transform taken not only at the scale considered but also across all finer scales available~\cite{muzyetal91}. 
  
      \item  In 1997 a functional analysis interpretation of the  scaling function for $q >0$ (see \eqref{foncech} below) opened the way to  determining when data can be modelled by locally bounded functions, and also to the first  mathematical results concerning the validity of the multifractal formalism for functions \cite{jmf2}.  
\end{itemize}

\subsection{Multiscale quantities and wavelet expansions} 

Definition  \eqref{mesreg} plays a key role in the derivation of the upper bound supplied by the multifractal formalism in the  measure setting, where the   scaling function $\eta_\mu (q)$ of a measure $\mu$ can be  defined as follows.   We will use the following notations for dyadic intervals:
\[\lambda \; ( = \lambda_{j,k} )  = \displaystyle \left[ \frac{k}{ 2^j},  \frac{k+1 }{ 2^j}\right[  \quad \mbox{ and }  \quad  3 \lambda = \displaystyle \left[ \frac{k-1}{ 2^j},  \frac{k+2 }{ 2^j}\right[.\] 
The { \em measure scaling function} of $\mu$ is defined by
\BE \label{foncdech}
\forall q \in \RR , \mbox{ if } \;\; S_\mu (j,q) =  2^{-j} \displaystyle\sum_{ k }  \left( \mu \left(3 \la_{j,k}  \right) \right) ^q, \quad 
\eta_\mu (q) =   \displaystyle\liminf_{j \rightarrow + \infty} \;\; \frac{\log \left(S_\mu (j,q)  \right)}{\log (2^{-j})}.
\EE
Denote by $\lambda_j (x) $ the unique dyadic interval of width $2^{-j}$  which contains $x$. Then   \eqref{mesreg}  can be rewritten as:
\BE\label{derviexp}
h_\mu (x)  =  \displaystyle\liminf_{j \rightarrow + \infty} \;\; \frac{\log \left(  \mu \left( 3 \lambda_j (x)  \right)   \right)}{\log (2^{-j})}.
\EE
When such a relationship holds between a non-negative quantity defined on dyadic intervals  and a pointwise regularity exponent, we will say that  the multiscale quantity (here $ \mu \left( 3 \lambda   \right)$) is { \em associated } with the corresponding exponent  (here $ h_\mu $). This notion is important because, when it holds, it follows that (see \cite{Jaffard2004})
\BE \label{formfetpmaj}
{\mathcal D}_f(H) \leq  \inf_{q }  \left( 1+Hq - \zeta_f (q)  \right).
\EE 

This created a strong motivation for the quest of multiscale quantities associated with pointwise regularity exponents. 
In the case of the H\"older exponent of a function,  a first possibility is to consider the  {\em first order oscillations} of $f$ 
\[ {\mathcal O}_f (\la)  =  \sup_{ x, y  \in\, 3 \la } |f(x) - f(y)| \]
(or higher order differences if H\"older exponents larger than one can be met in the data), see \cite{barral2011estimation,JAFFARD:2006:A}. However, this method does not present the (numerical and theoretical) advantages of wavelet-based methods (see \cite{Oscill} for recent  results  on this method). The state-of-the-art method makes use of {\em wavelet leaders} instead, which are defined as follows. 

Let $\psi$ be an oscillating and well localized function, with $r$ ($\geq 1$) first vanishing moments and of class $C^{r-1}(\RR)$. The function $\psi$ generates an $r$-{\em smooth orthonormal  wavelet basis}   when the $\{\psi_{j,k}(x)= 2^{j/2}\psi(2^{j}x-k)\}_{(j,k)\in \ZZ^2}$ form an orthonormal basis of $L^2 (\RR)$. The discrete wavelet coefficients of a function $f$ are defined by
\BE \label{defwc}
c_{j,k} =2^{j} \int_{\RR} f(x) \psi \left (2^j x-k \right)dx, \qquad (j,k)\in\ZZ^2.
\EE
For convenience, we will sometimes also index wavelets and wavelet coefficients by dyadic intervals and write indifferently $ c_\la = c_{j,k} $.

A first advantage of using wavelets is that they offer a numerically reliable extension of the Kolmogorov scaling function with a wider range of applicability: The {\em wavelet scaling function},  is defined as in \eqref{foncdech}, but replacing the multiscale quantity $\mu (3 \la_{j,k})$ by the wavelet coefficients $c_{j,k}$.   Let $q >0$; the { \em structure functions of order $q$} of  $f$ are defined as
\[\forall j \geq 0, \qquad  S^w_f (j,q) =  2^{-j} \displaystyle\sum_{ k }  \left| c_{j,k} \right|^q ;\]
and the {\em wavelet scaling function} of $f$ reads
\BE \label{wavscal} 
\zeta_f (q) =   \displaystyle\liminf_{j \rightarrow + \infty} \;\; \frac{\log \left(S^w_\mu (j,q)  \right)}{\log (2^{-j})} .
\EE

The definition of the wavelet scaling function does not rely on any assumption on the data (provided that the wavelet used is smoother than the maximal regularity encountered in the data), in which case \eqref{defwc} is interpreted  as a duality product between smooth functions (wavelets) and  a tempered distribution $f$. Furthermore, it is independent of the (smooth enough) wavelet basis which is used, see \cite{jmf2}.  
The scaling function thus obtained is still denoted  by  $\zeta_f (q)$ because it coincides with the Kolmogorov scaling function if  $q >1$ and if the H\"older exponent of $f$ takes only values below 1.  It has many use cases:
\begin{itemize}
  \item it can be used for classification; 
  \item it allows us to determine for which type of pointwise exponents a multifractal analysis can be performed, see  \eqref{foncech} below;
  \item it yields an upper bound of the increasing part of the {\em  weak scaling spectrum}, see Def. \ref{defwse} and Prop. \ref{majspecprop} below. 
\end{itemize}

The wavelet scaling function is not well-defined for negative $q$s; indeed, the distribution of wavelet coefficients for real-world data usually display a non-vanishing density around 0, hence wavelet coefficients of arbitrarily small size show up, and negative moments won't exist. 
A way  to mitigate this problem consists in considering the following alternative scaling function, see \cite{Jaffard2004}.

\BD 
Let  $f\in L^{\infty}_{loc} (\RR)$, and assume that a sufficiently smooth wavelet basis has been chosen. The wavelet leaders of $f$ are
$$\forall (j,k)\in\ZZ^2, \ l_{f}(\la )=
\sup_{ \lambda'\subset 3\lambda} \{|c_{\la'}|\}.$$
The wavelet leader scaling function is defined as
\BE \label{leadscal}
\forall q \in \RR , \mbox{ if } \;\; S_f (j,q) =  2^{-j} \displaystyle\sum_{ k }  \left( l_f (j,k) \right) ^q,  \mbox{ then} \qquad 
\eta_f (q) =   \displaystyle\liminf_{j \rightarrow + \infty} \;\; \frac{\log \left(S_\mu (j,q)  \right)}{\log (2^{-j})} . \EE
\ED 

A numerical advantage of using wavelet leaders in the definition of the scaling function is that, in contradistinction with wavelet coefficients,  their distribution vanishes around 0, see \cite{JAFFARD:2006:A,Lashermes08,Wendt07,Wendt2008c}, and  \cite{Wejdeneetal} for estimates of the laws of wavelet leaders and the application of statistical tests to derive empirical laws. An additional key  property is that, under a uniform regularity hypothesis on the data, wavelet leaders are {\em associated}  with  the H\"older exponent according to \eqref{derviexp}, see \cite{Jaffard2004}; it follows that the corresponding Legendre spectrum (using the wavelet leader scaling function)  yields an upper bound for the multifractal spectrum which holds without additional assumption. 

However, large classes of  signals cannot be modelled by locally bounded functions: In order to determine when this is possible, one computes the value taken by the { \em uniform H\"older exponent}  $\Hmin $, which is defined through a log-log plot regression 
\BE\label{defhmin}
\Hmin = \limsup_{j \to +\infty} \left( \frac{\log \left( \sup_k |c_{j,k}| \right)}{\log (2^{-j})} \right).
\EE 
This exponent  has found an independent interest for classification, see e.g. \cite{MandMemor} and Sec. \ref{secmegsig}.
If $\Hmin <0$, then $f$ is not locally bounded, see \cite{jmf2}, and it follows that a multifractal analysis based on the H\"older exponent cannot be performed; this situation is illustrated in Fig. \ref{figRMSEHmin}, on a MEG signal, yielding a negative value for $\Hmin$; see also Fig. \ref{figRMSEHmin} where it is shown that most exponents $\Hmin $ which we have estimated from MEG data are actually negative. 
In such situations, one  can  preprocess data by performing a { \em fractional integration}  of order $s$
prior to conducting multifractal analysis. 

Let us  briefly recall the definition of  fractional integration together with  the variant used in practice.
A function $f$ belongs to the { \em  Schwartz class}  if it belongs to $C^\infty$ and if all its derivatives have fast decay. The dual space of  the  Schwartz class is  the set of 
{ \em  tempered distributions}.  One  advantage of using this very general setting is that the  Fourier transform is well defined  on this space where it is a  one to one  mapping. This allows to define the fractional integral  or arbitrary order of a tempered distributions as follows:
\BE \forall s \in \RR, \qquad \widehat{f^{(-s)}} (\xi ) = (1+ | \xi|^2)^{-s/2}  \widehat{f} (\xi ).  \EE 
On the practical side, tempered distributions yield a general framework for modeling which requires no  a priori assumption on the data. 
In applications involving a wavelet analysis,  one does not compute explicitly the wavelet coefficients of $f^{(-s)}$, and one rather defines the {\em  pseudo-fractional integral} of order $s$, $\tilde{f}^{(-s)}$, by its wavelet coefficients as 
\[ \tilde{c}_{j,k}^{-s} = 2^{-sj} {c}_{j,k}. \] 
The function space interpretation of $H^{min}_{f}$ implies that
\[\forall s \in \RR, \qquad  H^{min}_{f^{(-s)}} = H^{min}_{\tilde{f}^{(-s)}} = H^{min}_{f}  + s ,  \] 
so that  it suffices to take a (pseudo-)fractional integral of order larger than $-H^{min}_{f} $ to  make the multifractal analysis based on the H\"older exponent  possible.  
The pointwise H\"older exponents of  $\tilde{f}^{(-s)}$  and ${f}^{(-s)}$  together with their different scaling functions coincide, see \cite{Wendt2009b}, 
which explains the choice of  using the {pseudo-fractional integral} in applications  (rather than the  fractional integral) since it  does not involve any additional computation.
It follows 
 that a multifractal analysis of ${f}^{(-s)}$  based on the H\"older exponent  can be carried out. 
  This technique  has often been used (either explicitly or implicitly) in multifractal analysis. In the continuous wavelet transform setting, it is for instance a prerequisite before using the WTMM (Wavelet Transform Modulus Maxima) method \cite{muzyetal91}, indeed, the continuous wavelet transform restricted at  its local maxima may yield  unbounded  quantities  if $H_f^{min} <0$.
Nonetheless, a (pseudo-)fractional integration can alter the shape of the multifractal spectrum in a way that cannot be a priori predicted, so that  it yields little  information on the initial data. 
This has been documented in the case of { \em Lacunary Wavelet Series}, where, for a given $p$, the multifractal spectrum of $f^{(-s)}$ is a shifted { \em and dilated} version of the multifractal spectrum of $f$, see \cite{Porqu2017}. 
In Sec. \ref{mathmod} we will 
show other examples which illustrate this phenomenon, and therefore call for a direct  analysis of the data  without such a preprocessing. 

When $H_f^{min} <0$,   a   direct analysis of the data without performing first fractional integration  is possible if one uses  weaker notions of pointwise  regularity which do not require that the analyzed function is locally bounded.  A first possibility is when $f \in L^p_{loc}$ for a $p \geq 1$. In that case, the following extension of the H\"older exponent introduced by A. Calder\'on and A. Zygmund can  be used \cite{Calderon61}. 

\BD\label{defp}
Let $f\in L_{\text{loc}}^{p} (\RR)$ with $p\geq 1$. Let $x_0\in\RR$. A function $f$ belongs to $\TPA$ when there exist a polynomial $P_{f,x_0}$ of degree less than $\alpha$ and constants $C,R>0$ such that
\BE\label{defTp}
\forall r\in (0,R), \quad \left( \frac{1}{r} \int_{x_0-r}^{x_0+r} |f(x)-P_{f,x_0}(x-x_0)|^p dx \right)^{\frac{1}{p}} \leq C r^{\alpha}. \nonumber
\EE
The $p$-exponent of $f$ at $x_0$ is $\HFP =\sup\{\alpha : f\in \TPA\}$.
\ED

Appropriate multiresolution quantities associated with the $p$-exponent have been introduced in \cite{jaffard2016p,leonarduzzi2016p}. They are referred to as { \em $p$-leaders}, and the corresponding multifractal formalism  allows to estimate the corresponding { \em $p$-spectrum} ; it is currently used in signal and image processing, and even preferred  to the wavelet leader based multifractal formalism due to its improved statistical performances, see \cite{leonarduzzi2017finite} where it is shown that values of $p$ close to $p=2$ should be preferred. 
The choice $p <1$ (which requires to replace the spaces $L^p$ by the real Hardy spaces $H^p$ in   Def.~\ref{defp}) allows  to analyze some classes of  tempered distributions, for instance when they are  supported by fractal sets,  see \cite{JaffToul,Jaffard06a}. 

 A simple criterium  allows to   determine  under which condition data can be modelled by a function in  $ L_{\text{loc}}^{p} $:  it is the case if  $\zeta_f (p) >0$. This follows from the   following interpretation of the wavelet scaling function in terms of regularity in the class of Sobolev spaces.  
Let $L^{p,s}$ denote the Sobolev space of  distributions whose fractional derivative of order $s$ belongs to $L^p$;  then \BE \label{foncech}  \zeta_f (p) = p \cdot \sup \{ s: \quad f \in L^{p,s}   \} \vspace{-2mm} ,\EE
  see \cite{jmf2}.   It follows that,  if $\zeta_f (p) >0$, then $f \in L^{p,s}$ for an $s >0$, so that $f \in L^p$. 

If  $\zeta_f (p) \leq 0$, then  one can still have recourse to a (pseudo-)fractional integration  in order to  estimate the  $p$-spectrum of a fractional integral of $f$.
Indeed if $s$ is large enough, then the wavelet scaling function of the smoothed signal $f^{-s}$ thus obtained  becomes positive for some  values of $p$: more precisely,  the Sobolev interpretation of the wavelet scaling function supplied by \eqref{foncech} together with the implication 
\[ f \in L^{p,t} \Longrightarrow f^{(-s)}  \in L^{p,t +s } \] (which follows directly from the definition of Sobolev spaces) implies that \[ \forall p >0, \qquad \zeta_{f^{(-s)}} (p) = \zeta_{f} (p) + ps; \] 
therefore, 
it suffices to take a fractional integral of order $s > {-\zeta_f (p)}/{p}$ to insure that $f^{(-s)}  \in L_{\text{loc}}^{p} $.

However, this  option, which requires the use of a fractional integral, meets the same limitations  as mentioned before: $p$-spectra can also be modified in  an unpredictable way under fractional integration. This drawback calls for the use of a multifractal analysis based on  another pointwise exponent which would be defined without any a priori assumption, i.e. in the very general setting supplied by tempered distributions, so that the corresponding multifractal analysis could  be directly applied to the data (and not to their fractional integrals), without restrictions.

\subsection{ The weak-scaling exponent} 

\begin{figure}[htbp]
    \centering
    \includegraphics[scale=0.7]{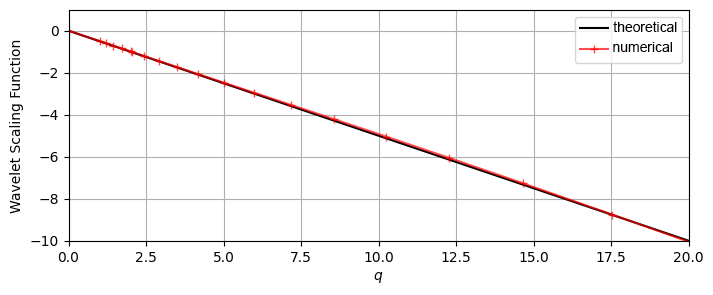}
    \caption{Numerical estimation of the wavelets scaling function for a Gaussian white noise (in red) superimposed on the theoretical result (in black)}
    \label{figwavscal}
\end{figure}

 In practice, the use of $p$-exponents does not cover all types of data that are met in real-world applications. This has been noticed for the analysis of the cadence of marathon runners, for instance, some of which  verify that $\forall p >0$, $\eta_f (p) <0 $  \cite{Farf4};  and it also happens when analyzing brain activity notably on MEG signals, as shown later in Sec. \ref{secmegsig}.  On the theoretical side, a  simple example  for which no $p$-exponent can be used is supplied by Gaussian white noise; indeed, the fact that its  coefficients on any orthonormal basis are IID centered normal Gaussian variables easily implies that its wavelet scaling function is 
 $$ \forall q>0, \qquad \zeta_X (q) = -\frac{q}{2},$$ so that it takes  negative values only (see Fig. \ref{figwavscal}). 
 
 We will see in Sec. \ref{mathmod}  other examples of mathematical models for which no $p$-exponent can be used. These situations, which show up both in theory and  applications, call for the use of another pointwise regularity exponent which would be defined without  any a priori assumption. Such an exponent has  been introduced by Y. Meyer in \cite{MeyWVS}, with a different  purpose. The initial motivation  was to answer a problem which appeared in the mid 1980s: Indeed, it was commonly believed that the pointwise H\"older exponent of a function $f$ can be  characterized by the decay rate of its continuous wavelet transform in the {\em cone of influence} of the point considered; if translated to the discrete wavelet setting, this means that, for a given point $x_0$ 
 \BE  \label{twomicro} \exists C, C' >0: \quad \mbox{ if }\quad  \left| \frac{k}{2^j} -x_0 \right| \leq   \frac{C}{2^j} \quad \mbox{ then }  \qquad | c_{j,k} | \sim 2^{-h_f (x_0)  j}  \EE
 (the $2[C] +1$  wavelet coefficients closest  to $x_0$ at each scale decay like $2^{-h_f (x_0)  j}$). 
 Such a statement was proved wrong, typical counterexamples being supplied by the { \em chirps } 
\BE \label{chirp}  x \rightarrow | x-x_0|^{\alpha} \sin \left( \frac{1}{| x-x_0|^{\beta}}\right), \EE for $\alpha, \beta >0$.  Yves Meyer made a precise analysis of how \eqref{twomicro}   can be interpreted, and he showed that it can be associated with a new pointwise regularity exponents. 
To state his results, we need to recall the following notion, which was introduced by J.-M. Bony \cite{bony1984two}, and used as a key tool in the wavelet characterization of pointwise regularity \cite{jaffard1991pointwise}.

\BD \label{defmicro} 
A tempered distribution $f: \RR \rightarrow \RR$ belongs to the two-microlocal space $C^{s,s'} (x_0)$
if  its wavelet coefficients  (in an $r$-smooth wavelet basis with $r > \max (|s|, |s'|)$) satisfy
\begin{equation} \label{deumic}  \exists C, \;\; \forall j,k,  \qquad | c_{j,k} | \leq C 2^{-sj} (1+ | 2^j x_0 -k|)^{-s'} .
\end{equation}
\ED 

This definition is independent of the wavelet basis used, see \cite{jaffard1991pointwise}. 
 Yves Meyer  introduced the following notions in  \cite{MeyWVS}.

\BD \label{defwse} 
A tempered distribution $f: \RR \rightarrow \RR$  belongs to 
 $  \Gamma^s (x_0)$
if   there exists $s' >0$ such that $f\in C^{s,-s'} (x_0)$. 

The weak-scaling exponent of $f$  at $x_0$ is 
\[ h^{ws}_f (x_0)= \sup \{ s : \;  f \in \Gamma^s (x_0)\}.\]
\ED

 Yves Meyer  showed that this definition allows to give a precise mathematical content to the (loose) statement \eqref{twomicro}.

 The multifractal weak scaling spectrum ${ \mathcal D}^{ws}_f : \RR \cup \{ + \infty \}  \rightarrow \RR_+\cup \{-\infty\}$ of $f$ is the mapping defined  by
$$ \forall H \in \RR, \qquad 
{ \mathcal D}^{ws}_f (H) = \dim_H \left( \{x\in\RR : \quad h^{ws}_f (x)=H\} \right)  , 
$$
see  \cite{Farf4} (where  equivalent definitions of the weak scaling exponent are also derived).

A direct consequence of Def. \ref{defmicro} is that, for any distribution $f$, 
 \begin{equation} \label{www1} 
     \forall x, \qquad  h^{ws}_{f'}  (x)  = h^{ws}_f (x)  -1 , 
 \end{equation}
 so that 
 \begin{equation} \label{www2}   \forall H \in \RR, \qquad 
{ \mathcal D}^{ws}_{f'}  (H) = { \mathcal D}^{ws}_{f}  (H+1)   .  
\end{equation}

Weak scaling exponents can take any positive or negative value. In particular, this notion allows us to give a proper mathematical framework for defining pointwise singularities of arbitrary negative exponent. This is not a straightforward problem: for instance, it is well known that the usual {\em cusp singularites}
\[ x \rightarrow | x-x_0|^\alpha \] 
no longer make sense if $\alpha <-1$; indeed, they are ill-defined as Schwartz distributions, so that, for instance, their wavelet coefficients cannot be properly defined, no matter how smooth the wavelet used is.\par

In this article, the purpose is to investigate techniques for the estimation of the function ${ \mathcal D}^{ws}_f $, and to  show its relevance for  the analysis of MEG data. 

In   Sec. \ref{sec21},  we discuss the  limitations of the use of  $p$-exponents  to perform  multifractal analysis.  In Sec. \ref{uppbousec} we  show how the increasing  part of the  weak scaling spectrum can be estimated directly from  wavelet coefficients. The estimation of the decreasing part requires the use of  
  $(\theta, \omega)$-leaders as  multiscale quantities introduced in in Sec. \ref{secdeflead}. 
In Section \ref{mathmod}, we illustrate the use of these multiresolution quantities  by showing what the corresponding multifractal  analysis yields for several classical mathematical models, such as fractional Gaussian noises, random wavelet series and multifractal random walks, thereby demonstrating the relevance and accuracy of the weak-scaling spectrum compared to previously introduced methods. Finally, in Sec. \ref{secmegsig}, we apply this technique to MEG recordings~(time series), for which a multifractal analysis based on $p$-exponents cannot be systematically completed. 

This article is partly review and partly research: Besides the introduction, the review part concerns Section \ref{sec2}, where we collect several results  concerning the weak-scaling exponent  scattered in the literature, and  complement them by new results. Sections  \ref{mathmod} and 
\ref{secmegsig}  contain new material.

 \section{Mathematical tools for weak-scaling multifractal analysis}

\label{sec2}

In this section, we collect  results concerning  equivalent mathematical definitions of the weak-scaling exponent  and its relevance for multifractal analysis.  Furthermore, we discuss how this analysis can be performed in a numerically stable and tractable way.  However,  we start by delimiting the situations where no other exponents can be used; this is the purpose of Sec. \ref{sec21} where we discuss criteria under which the $p$-exponent can be used. 

\subsection{Limitations for the use of the \texorpdfstring{$p$}{p}-exponent}

\label{sec21}

Several criteria have been proposed to determine if a multifractal analysis based on the $p$-exponent can be worked out. A simple  one already mentioned is that  it can be used  when
 $\zeta_f (p) >0$,
see \cite{jaffard2016p}.  Another criterion, which is derived from the {\em large deviation spectrum} of the wavelet coefficients, can be found in \cite{Farf4}. We now propose a new one that can be applied  when  some information is available concerning the location of the singularities of the data.  

\BD \label{defsplit} Let $\delta < 1$ and $q >0$. A tempered distribution $f: \RR \rightarrow \RR$ is $(\delta, q)$-sparse
if it  can be written $f = f_1 + f_2$ with  $f_1 \in L^q$ (or, when $q \leq 1$,  if $f_1$ belongs to the Hardy real space  $H^q $ ) and the wavelet 
expansion  of $f_2$ in a given wavelet basis is such that, at generation $j$,  $f_2$
has at most $C \cdot  2^{\delta j}$ nonvanishing wavelet coefficients.
\ED  

Typical example of  $(\delta, q)$-sparse distributions are provided by lacunary wavelet series~\cite{Jaf6} or by distributions supported by a fractal set of upper box dimension $\delta < 1$.  This last case is relevant e.g. for applications in urban modeling, where data are carried by the urban network, which is often modelled by a fractal set \cite{frankhauser1994fractalite}. 

Recall that the Besov space $B^{s , \infty}_{p} $ can be characterized by the following wavelet condition:
\[ \exists C, \; \forall j , \qquad  2^{-j} \sum_k | c_{j,k} |^p  \leq C 2^{ -spj} .  \] 

\BP \label{splitting} If $f$ is a  $(\delta, q)$-sparse distribution, then  there exist $\ep, p >0$  such that  $f \in B^{\ep , \infty}_{p}  $, so that  a multifractal analysis of $f$ using $p$-exponents can be performed. 
\EP

{ \bf Proof: }  Since $f_1 \in L^q$,  a $p$-exponent based multifractal analysis  of  $f_1$ can be performed   for any $p \leq q$, so that we focus on $f_2$. Since it is a tempered distribution, it is of finite order, so that there exists $A \in \RR$ such that 
$f_2\in C^A (\RR) $ (one can pick any $A < H_f^{min} $);  thus, its wavelet coefficients satisfy 
\[ \exists C \quad \forall j, k , \qquad  | c_{j,k} | \leq C 2^{-Aj}.  \]
If $A >0$, then $f_2$ has a positive uniform H\"older regularity, and the result holds. Let us now assume that $A \leq 0$. 
Since   $f_2$
has at most $C 2^{\delta j}$ nonvanishing wavelet coefficients,
\[ \forall j \qquad  2^{-j} \sum_k | c_{j,k} |^p  \leq C 2^{(-1 + \delta) j} 2^{-Apj},  \]
so that \[  f_2 \in  B^{s , \infty}_{p} \quad  \mbox{ for  any } \quad   s \leq  \frac{1-\delta   }{p} +A.   \] 
We now recall the following classical embeddings between Besov and Sobolev spaces
\[ \forall s > s' > s", \qquad B^{s , \infty}_{p} \subset L^{p,s'} \subset  B^{s" , \infty}_{p}  .    \] 
It follows that $p$-exponents can be used as soon as $f \in B^{s , \infty}_{p}  $ for an $s>0$,
and $s$ can be picked positive as soon as 
\BE \label{valp}  p < \frac{1-\delta}{- H^{min}_f} .\EE

In order to be useful in applications,  this criterium requires to estimate the value of $\delta$. Its value can be  known beforehand for some types of data (such as, in 2D, for urban data which are carried by a fractal set the box dimension of which can be estimated). In other applications,  one has to construct a  practical algorithm for the determination of the splitting  supplied by Def. \ref{defsplit}.  We  now propose one which is based on a splitting of the wavelet coefficients of $f$.

Let $C$ be a fixed constant which is related to the $L^q$ norm of  the function $f_1$ which will be constructed. 
At each generation $j$, we  denote by  $S_{ N, j} $ the restriction of the quantity 
\[ \sum_{k} | \cjk |^q   \] 
to the $N$ smallest values taken by  $| \cjk |$. This is clearly an increasing function of $N$  and we denote by $N_j$ the largest value of $N$ such that 
\[  S_{ N, j}  \leq C \frac{2^j}{j^{2q}} . \]
This defines  a splitting  of the wavelet coefficients of $f$ into two sets, and we denote by  $K_j$ the set of ``small'' coefficients thus  selected, and the  complementary set by $L_j$. We then define 
\[ f_1 = \sum_j \sum_{ k \in K_j } \cjk \pjk  \quad \mbox{ and }  \quad f_2 = \sum_j \sum_{ k \in L_j } \cjk \pjk.  \]
By construction,  $f_1 \in B^{0, q}_q \subset L^q$.  Let $M_j$ be the total number of wavelet coefficients  computed at generation $j$. The number of nonvanishing coefficients of $f_2$ is $M_j -N_j$, so that $\delta$ can now be derived through a log-log plot regression:
\[ \delta = \limsup_{ j \rightarrow + \infty}  \frac{ \log (M_j -N_j ) }{ \log (2^j) } . \] 


\subsection{An upper bound for the weak-scaling spectrum}

\label{uppbousec}

The  wavelet scaling function  allows to derive  the following  upper bound of  the  weak-scaling spectrum see  \cite{JaffMel}.
  
  \BP \label{majspecprop} Let $f$ be a tempered distribution defined on $\RR$. Then  its weak scaling spectrum satisfies 
\begin{equation} \label{theo11} D^{ws}_f (H) \leq \inf_{ p >0  } \left( Hp -\eta_f (p)  +1\right) . \end{equation}
\EP  

This result is natural if we remember  the heuristic which motivated the  introduction of the weak-scaling exponent: The size of  the wavelet coefficients located in the ``cone of influence'' yields this exponent through a log-log plot regression; this means that, though   wavelet coefficients are not multiscale quantities associated with the weak-scaling exponent in the sense  supplied by \eqref{derviexp}, nonetheless they  are ``close'' to be such, so that the corresponding upper bound still holds.  
However, this formula meets  a severe  limitation: 
 Since the infimum is taken on positive $p$s only, the  right hand side of \eqref{theo11} is increasing, and this bound can only estimate the increasing part of the spectrum. 
Actually, if this formula is applied to negative $p$s, it  does not yield a sharp estimate,  as shown by  the toy-example  supplied by Brownian motion, see \cite{JAFFARD:2006:A,Farf4}.

An extension of Prop. \ref{majspecprop} for $p <0$ is proposed in \cite{JAFFARD:2006:A}. It is based on structure functions which are not derived  directly from wavelet coefficients, but rather from { \sl $\ep$-leaders}, i.e. from multiscale quantities which are defined as local suprema of wavelet coefficients taken on small boxes of width $2^{\ep j} $ around the corresponding location of the wavelet  coefficients in the time-scale half-plane, and then taking a limit  of the resulting scaling functions  when $\ep \rightarrow 0$.  This formulation however is not fitted to  applications, because of  the double limit  which is involved in this approach. This motivated the  introduction of  new multiscale quantities which we now describe; indeed, they do not present this double-limit drawback and  they yield sharp upper bounds for the weak scaling  spectrum (see Def. \ref{defwse}  below), which turn out to be equalities for several classes of models, see \cite{Farf4} and Sec. \ref{mathmod}. 

\subsection{Multiscale quantities: \texorpdfstring{$(\theta, \ome)$}{(theta,omega)}-leaders }

\label{secdeflead}

We now define the local suprema of wavelet coefficients as  multiscale quantities on which multifractal analysis for the weak-scaling exponent will  be based.

A function 
$\theta:  \NN \rightarrow \RR^+$ has { \em sub-polynomial growth}  if it satisfies
\BE \label{condome2} 
\left\{ 
\begin{array}{l}
 \forall j  , \quad     \theta (j+1) \geq \theta (j) \geq j  \\
       \\
  \displaystyle\limsup_{ j \rightarrow + \infty} \displaystyle\frac{\log (\theta (j)  -j )}{\log (j)} <1  ; 
\end{array}
\right.
\EE
typical examples are supplied by  functions  of logarithmic growth $ \theta (j) =  j +   C (\log j)^\alpha$ for an $\alpha \geq 0$, or by power laws $\theta (j) = j+ j^\beta  $ for a $\beta <1$.  Note that this definition is slightly more general than those introduced previously, see \cite{Farf4} and references therein.

   A function 
$\ome:  \NN \rightarrow \RR^+$ has { \em sub-exponential growth} if it is   non-decreasing  and  such that 
\BE \label{condome} 
\left\{ 
\begin{array}{ccc}
  \ome (j) \rightarrow + \infty &  \mbox{ when}  &  j \rightarrow + \infty \\
  &   &   \\
  \displaystyle\frac{\log (\ome (j) )}{j} \rightarrow 0 &   \mbox{ when}  &  j \rightarrow + \infty   ; 
\end{array}
\right.
\EE
typical examples are supplied by power-laws $j \rightarrow  j^a$ for an $a >0$. 

\begin{figure}[htbp]
    \centering
    \includegraphics[scale=0.2]{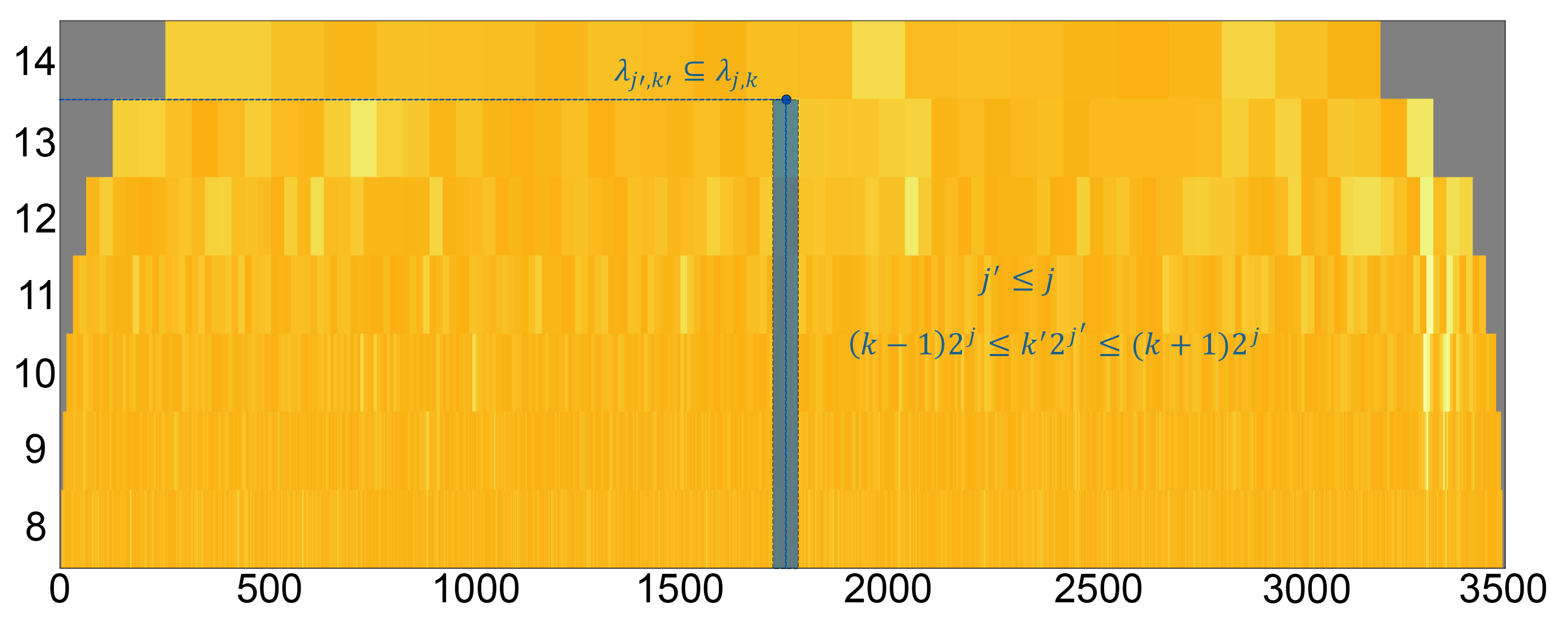}
    \includegraphics[scale=0.2]{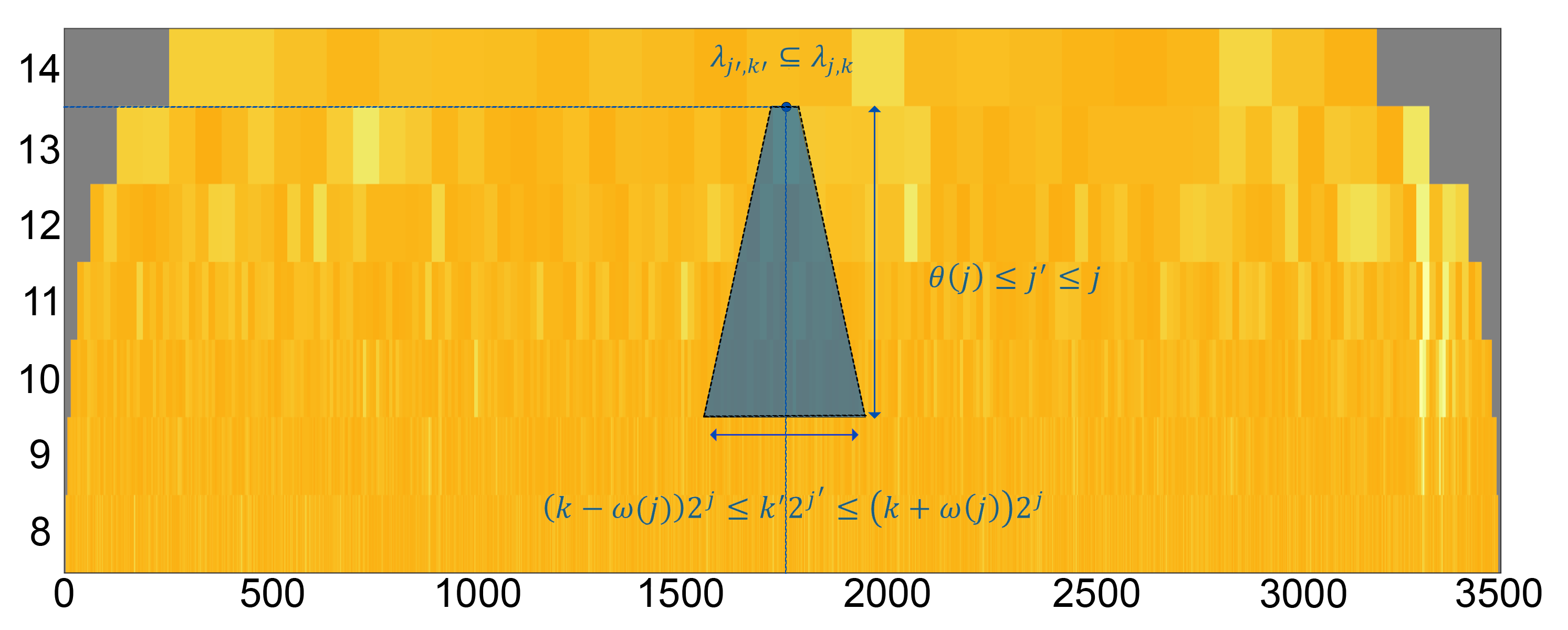}
    \caption{Selected wavelet coefficients for leaders (top) and $(\theta,\omega)$-leaders (bottom) for the determination of the corresponding multiresolution quantities. }
    \label{figselect}
\end{figure}

We start by defining the sets of dyadic intervals  on which the local suprema of wavelet coefficients will be taken. 

\BD  \label{def:thetaomeganeigh} Let  $\theta$ and $  \ome $ be two functions   with respectively sub-polynomial and  sub-exponential growth, and let $\lambda \;  ( = \lambda_{j,k} ) $ be a dyadic interval;   the $(\theta, \ome)$-neighbourhood of $\la$, denoted by $V_{ (\theta, \ome)}  (\la)$  is the set of dyadic intervals $\la' \;  ( = \lambda_{j',k'} ) $,  indexed by the  couples $(j', k')$ satisfying 
 \[ j \leq j' \leq  \theta (j)  \quad  \mbox{ and } \quad \left| \frac{k}{2^j} -\frac{k'}{2^{j'}} \right| \leq  \frac{\ome (j')}{2^j} . \] 
 \ED
 
 We now introduce  the $(\theta, \ome) $-leaders which will be the multiscale quantities on which the multifractal analysis of the weak-scaling exponent will be based, see Figure \ref{figselect}. 
 
 \BD \label{def:thetaomega} 
  Let $f$ be a tempered distribution of wavelet coefficients $(c_{j,k})$; 
 the $(\theta, \ome) $-leaders  of $f$ are defined by 
 \BE \label{defomelead} d_{j,k} = \sup_{(j', k')  \in V_{ (\theta, \ome)}  (j,k) } | c_{j',k'} | . \EE   
\ED

This definition is slightly more general than the one proposed in \cite{Farf4}; its motivation is to answer  numerical problems met by the previous definition, while keeping its key mathematical properties. Indeed, it allows for extra flexibility in the  choice of the number of wavelet coefficients  on which  the supremum  is taken: Note that, with this definition, this supremum  is taken on  $2 \omega (j) + O(1) $ at the generation  $j$, $4 \omega (j) + O(1) $ at  the generation $ j+1$, ... , so that, adding up, it is taken on 
\[ \left( 2^{ \theta (j) +1} -1\right) \omega (j)  + O(j)  \]
coefficients.  

The definition of $(\theta, \ome) $-leaders yields an  extension of  the wavelet scaling function \eqref{wavscal} to  $p<0$; indeed, one can easily check that, for $p >0$, the following definition coincides with \eqref{wavscal}. 
 
\BD Let $f$ be a tempered distribution; its  wavelet scaling function    is  defined by 
  \BE \label{defscalond2}  \forall p \in \RR , \qquad
  \zeta_f (p) =   \displaystyle \liminf_{j \rightarrow + \infty} \;\; \frac{\log \left( \omega (j) \cdot 2^{-j } \displaystyle\sum_{ k= l \cdot [2\cdot {\ome (j)}] }  | \djk |^p  \right) }{\log (2^{-j})}. \EE
  \ED 
  
 The sum is taken over the multiples of $[2\cdot {\omega (j)} ] $ so that the contribution of one dyadic interval $\lambda'$ is taken into account only once, inside one of the $(\theta, \ome) $-leaders.  If the wavelet coefficients are computed over an interval of length $L$  then,  at the generation $j$, there are  $\sim L 2^j $ wavelet coefficients which are computed; since the supremum in the computation of wavelet leaders  is taken on $[2 \cdot \omega (j)]  +1 $ coefficients of generation $j$, then   the prefactor of normalization of the sum in \eqref{defscalond2} corresponds to the number of elements on which this sum is taken.

The following result  was already derived in \cite{JAFFARD:2006:A} in the case of $\ep$-leaders
and extended to  $(\theta, \omega)$-leaders in \cite{Farf4}. One easily checks that it remains valid for the extension of $(\theta, \omega)$-leaders that we propose in the present paper. 
 
 \BP \label{majspecprop2} Let $f$ be a tempered distribution. Then its weak scaling spectrum satisfies 
\begin{equation} \label{theo112} \dws (H) \leq \inf_{ p\in \RR   } \left( Hp -\zeta_f (p)  +d\right) . \end{equation}

In particular, if the wavelet scaling function of a distribution $f$ is a linear function over $\RR$, then its weak scaling exponent is constant. 
\EP

\section{Mathematical models}

\label{mathmod}

\subsection{  Fractional Gaussian noise } 
\label{secfbm} 

\begin{figure}[htbp]
    \centering
    \includegraphics[scale=0.6]{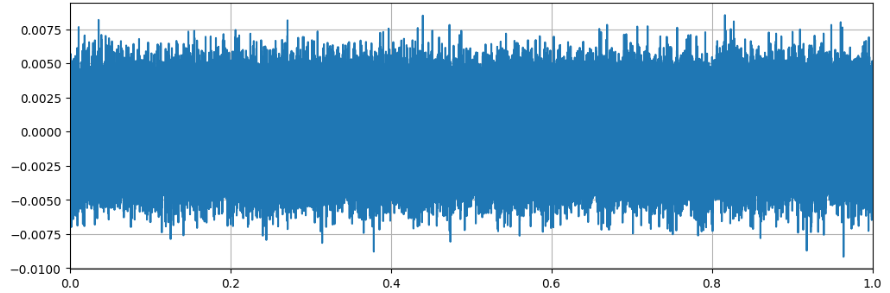}
    \caption{Simulation of a fractional Gaussian noise (fGn) $W_\alpha$ with Hurst exponent $\alpha=-0.5$. In this specific case, it is simply referred to as white Gaussian noise.}
    \label{fignoise}
\end{figure}

We start by  considering  fractional Gaussian noises (fGn) which play an important role in modelling (see Fig. \ref{fignoise}). Their sample paths are  random Schwartz distributions, which we denote by $W_\alpha  $; their Hurst exponent is negative: $\alpha \in (-1, 0)$ and their sample paths do not locally belong to any $L^p$ space,  so that their multifractal analysis cannot be performed using $p$-exponents. One way to recover this property consists in remarking that  
$W_\alpha  $ is a sample path by sample path derivative (in the sense of distributions) of  a fBm $B_{\alpha +1}$, whose Hurst exponent satisfies $\alpha +1 \in (0, 1)$. Since the wavelet scaling function of $B_{\alpha +1}$  is $\zeta_{B_{\alpha +1}} (q)= (\alpha+1)q$, it follows that
\[\forall q >0, \qquad  \zeta_{B_\alpha} (q)= \alpha q,  \]
hence always takes negative values. This justifies the use of the weak scaling exponent in order to analyze its pointwise regularity.

\begin{figure}[htbp]
    \centering
    \includegraphics[scale=0.8]{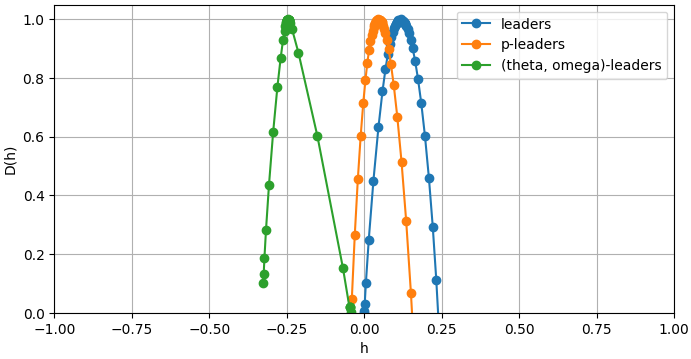}
    \caption{Legendre spectrum of a fGn with Hurst exponent $\alpha=-0.25$ obtained using three methods: leaders (in blue), $p$-leaders (in orange), and $(\theta,\omega)$-leaders (in green) with $\theta(j)=j+j^{0.25}$ and $\omega(j)=j$}
    \label{figspectnoise}
\end{figure}

The weak scaling exponent of  $B_{\alpha +1}$  satisfies 
\[ \mbox{ a. s. } \quad \forall x \in \RR,  \qquad  h^{ws}_{B_{\alpha +1}}   (x) = \alpha+1. \] 
It follows from \eqref{www1}  that  the  weak scaling exponent of $W_\alpha$  satisfies 
\[ \mbox{ a. s. } \quad \forall x \in \RR,  \qquad  h^{ws}_{W_\alpha}   (x) = \alpha ,  \] 
so that, using \eqref{www2},   its weak scaling spectrum is given by 
\[ \left\{ \begin{array}{rl} {\mathcal D}^{ws}_{W_\alpha}  (H )= & \gamma \mbox{ if }  H  = \alpha  \\  & \\ = &  -\infty \mbox{ else.} \end{array}  \right. \] 
We now inspect if this result can be recovered through a numerical estimation of the Legendre spectrum of $W_\alpha$.  Fig. \ref{figspectnoise} shows the Legendre spectrum  a fGn with Hurst exponent $\alpha=-0.25$.  Its Legendre spectrum is estimated  using wavelet  leaders, $p$-leaders,  and $(\theta,\omega)$-leaders.  The first two methods yield wrong results as expected, whereas the use of   $(\theta,\omega)$-leaders yields a Legendre spectrum  sharply peaked at the right value $\alpha = -0,25$.

\subsection { Multifractal analysis of random wavelet series } 

Random wavelet series (RWS) were  introduced in  \cite{AJ02} where their multifractal analysis was performed. They offer an interesting field of investigation in order to compare the different variants of multifractal analysis; indeed,   their multifractal spectra differ depending if one uses the H\"older and the $p$-exponents, and  it depends on the value of $p$ that is chosen, see Theo. \ref{theo:almosteverywhere} below.  We show in this section that a multifractal analysis based on the weak-scaling exponent yields  yet another spectrum, which supplies more information on the parameters  which characterize the RWS.   We start by briefly recalling the construction of these processes.

\begin{defi} \label{defrws} Let $(\pjk)_{j,k \in \ZZ^2}$ be a smooth orthonormal wavelet basis.  A RWS   associated with this basis   is a stochastic process of the form
\BE \label{defxt}  X_t =  \sum_{ j\geq 0} \sum_{ k\in \ZZ}  \cjk \pjk (t)  \EE 
such that  its   wavelet coefficients $\cjk$ 
are 
independent and, at each scale $j$,   share  a common law $\mu_j$. 
Additionally,  these laws satisfy 
\BE \label{hypcj} \mbox{ a.s. }\quad  \exists   C >0 ,  \;\;  \exists   A\in \RR   , \; \; \forall j \geq 0, \; \;\forall k \in \{ 0, \cdots 2^{j} \} ,  \qquad |\cjk  | \leq C 2^{- A j} . \EE
\end{defi}

Note that this notion is not canonical, but depends on the wavelet basis chosen. 
Since we are interested in  regularity properties of the sample paths of $X_t$, we need not care about possible terms corresponding to $j <0$  which would yield a smooth contribution to \eqref{defxt}, and we do not consider such a  component in the following. 
Note that the assumption \eqref{hypcj} only implies that the sample paths of the process are well defined as a Schwartz distribution; more precisely, it   implies  that  the process $X$ has some uniform regularity: the wavelet characterization of the H\"older spaces implies that   a.s. the sample paths of $ X $ locally belong to the H\"older space  $C^{A}_{loc} $. A simple sufficient condition implying that the sample paths are continuous (and thus that the H\"older exponent can be used in order to estimate pointwise regularity) is to  pick  $A>0$  in \eqref{hypcj}.  Another condition implying that the sample paths belong to $L^p_{loc}$ is given below, see Prop. \ref{lprws}.  

The a.s. multifractal properties  of the sample paths of RWS depend on a quantity called the \emph{wavelet large deviation spectrum} introduced in   \cite{AJ02}, and which we now recall.  Let $j \geq 0$ be given and denote by $\bsb{\rho_j}$ the common
probability measure of the $2^j$ random variables $ X_{j, k} \deq {-\log_2(|
  \cjk |)}/{j} $.  Thus  $\bsb{\rho_j}$  satisfies
\begin{equation*}
  \Proba\paren{\abs{\cjk} \geq 2^{- \alpha j}} = \bsb{\rho_j}( (- \infty,
  \alpha]).
\end{equation*}

\begin{defi} \label{defldws}  Let   $X_t$ be a RWS.  Let  
\begin{equation*} \forall \alpha \in \RR, \quad \mbox{let} \quad 
  \brho(\alpha, \ep) \deq \limsup_{j \rightarrow +\infty} \frac{
    \log_2 \left( 2^j \bsb{\rho_j} ( [\alpha -\ep, \alpha +\ep ])\right) }{j},
\end{equation*}
and, for $\alpha = +\infty$,  
\begin{equation*} 
  \brho(A) \deq \limsup_{j \rightarrow +\infty} \frac{
    \log_2 \left( 2^j \bsb{\rho_j} ( [A, +\infty) )\right) }{j},
\end{equation*}
The {wavelet large deviation spectrum}  of $X$ is 
\begin{equation}
  \label{eq:2}
 \mbox{ if } \alpha < +\infty, \mbox{ then } \qquad \brho(\alpha) \deq \inf_{\ep > 0} \brho(\alpha, \ep),
\end{equation}
\begin{equation}
 \label{eq:3}
 \mbox{ if } \alpha =  +\infty, \mbox{ then } \qquad \brho(+\infty ) \deq \inf_{A > 0} \brho(A).
\end{equation}
The support of the wavelet large deviation spectrum  is 
\[ supp (\brho) = \{ \alpha : \quad \brho (\alpha ) \geq 0\} . \]
\end{defi}

Note that $\brho$ is defined on $\RR \cup \{ + \infty\}$ and takes values in $ [-\infty , 1] $. 
As in \cite{AJ02}, in order to evacuate degenerate cases of little interest, we suppose that $\brho(\alpha)$ takes a positive value for
at least one (finite) value of $\alpha$.

The following result follows from the determination of the wavelet scaling function of RWS in \cite{AJ02}; it    supplies a sufficient condition  for the use of the  H\"older exponent or the $p$-exponent in the multifractal analysis of $X_t$.

\BP \label{lprws} Let $p \in (0, +\infty)$. 
If 
\BE \label{condlp} \forall \alpha \in \RR,  \qquad  \brho (\alpha ) < p \alpha +1    \EE
then the sample paths of $X_t$ almost surely belong to $L^p_{loc}$.
Furthermore,  if \[ \exists \ep >0: \quad \forall \alpha < \ep,   \qquad  \brho (\alpha ) = -\infty     \]
(or, equivalently, if \eqref{hypcj} holds  for an $A >0$), then the sample paths of $X_t$ almost surely belong to $L^\infty_{loc}$.
\EP

Let 
\BE W = \{ \al: \quad \forall \ep >0, \qquad  \sum_{j \in \NN}  2^j \brho_j([\alpha-\ep,\alpha+ \ep]  )=+\infty  \}, \quad 
  \bhmin \deq \inf_{\alpha } W,  \EE
and
\begin{equation*} 
  \bhmax (p)  \deq \paren{\sup_{\alpha } \frac{\brho(\alpha)}{\alpha + 1/p }}^{-1}.
\end{equation*}
The following result  yields the multifractal $p$-spectra of the  sample paths of  RWS. 
The case $p = + \infty$ corresponds to the H\"older exponent.

\BT
  \label{theo:almosteverywhere} 
  Let $X$ be a random wavelet series, and assume that \eqref{condlp} holds. With
  probability one, the sample paths of $X$ share the following properties:
  \begin{itemize}
  \item The  support of their multifractal $p$-spectrum is 
  $     { \cal S}_X =  [  \bhmin,  \bhmax (p) ] $;  
  \item their multifractal $p$-spectrum $ { \cal D}_X (H)$ is  given by     \begin{equation}
      \label{dhas} \forall H \in { \cal S}_X , \qquad 
      { \cal D}^p_X (H) = H\sup_{\alpha \leq H} \; \brho(\alpha) \frac{H+ 1/p}{\alpha + 1/p};
    \end{equation}   
     \item for almost every $t$,
    \begin{equation}
      \label{eq:hache}
      h_X(t) = \bhmax . 
    \end{equation}
    \item the Legendre $p$-spectrum is the concave hull of the multifarctal spectrum. 
  \end{itemize}
\ET

The last statement is a weak formulation of the multifractal formalism. 
 This theorem is  proved in  \cite{AJ02} in the case of the H\"older exponent and in \cite{Porqu2017} for the $p$-exponent in the case of lacunary wavelet series (i.e. when $\brho$ take only one non-negative value). Its extension to the general case of the $p$-exponent of RWS  follows from adapting the ideas developed in \cite{AJ02}  inside the framework supplied by  $p$-exponents as shown in \cite{Porqu2017}. 

We now consider the  setting supplied by the weak scaling exponent. 

\BT Let $X$ be a random wavelet series. 
The weak scaling mutifractal spectrum of    $X$  is given by 
  \[\forall H \in \RR , \mbox{ a.s.,  }  \qquad   { \cal D}^{ws}_X (H)= \brho (H)\; 1_W (H) .  \]
\ET

 Sketch of proof: This theorem  follows from several results of  \cite{AJ02}. Let $\ep >0$ 
 and denote by $E_\alpha^\ep$ 
 the limsup of  the $\ep$-neighbourhoods of  the dyadic intervals $\la$ 
 such that the corresponding wavelet coefficient  $\cjk$ satisfies  $\cjk \sim 2^{-\alpha j} $. 
 First, note that outside of the set 
 \[ \bigcup_{\alpha} E_\alpha^\ep   \]
 $h^{ws}_X $ takes the value $+\infty$.  Letting $\ep \rightarrow 0$, we obtain that the support of the spectrum is included in the support of $1_W $. 
 Let now $\alpha$ be fixed;  for any $\ep >0$, the set of points $x$ where $h^{ws}_X (x) =\alpha$ is included in 
 \[ F_\alpha^\ep =  E_\alpha^\ep  - \bigcup_{\beta \neq \alpha}  E_\beta^\ep ,  \]
 and  a simple box-counting argument yields that 
\[ \dim (F_\alpha^\ep )  \leq   \brho (\alpha ) + o(1) \]
(where the $o(1)$ has to understood as a limit when $\ep \rightarrow 0$).  Taking the limit when $\ep \rightarrow 0$, it follows that  ${ \cal D}^{ws}_X (H) \leq \brho (H)\; 1_W (H)$.  
The lower bound is obtained as in \cite{AJ02}, using an   ubiquity-type argument.

\subsection { Multifractal random walk (MRW)  }

\begin{figure}[htbp]
    \centering
    \includegraphics[scale=0.19]{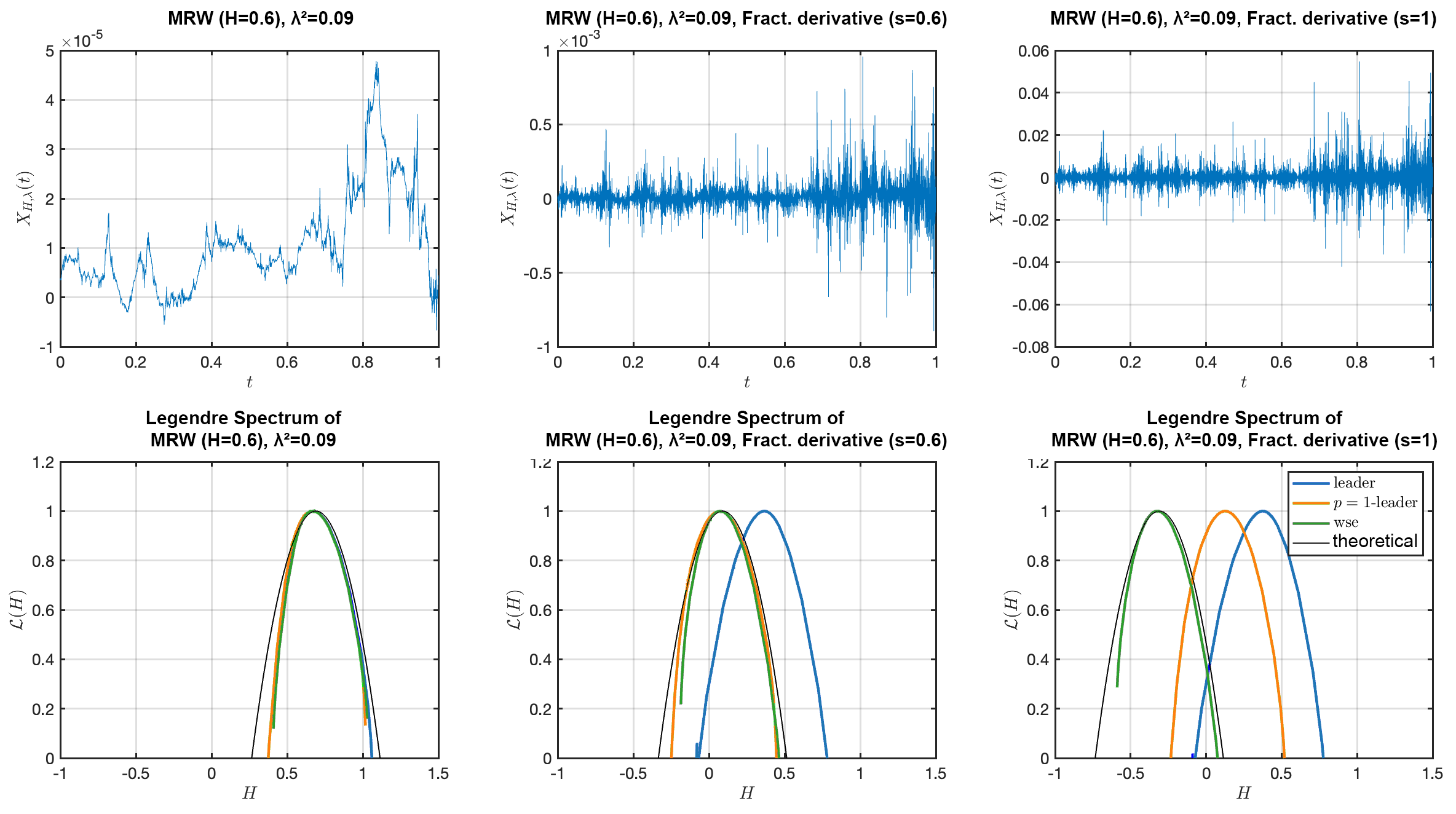}
    \caption{{\bf Compared multifractal analyses for Multifractal random Walks (MRW).} Top row, from left to right, MRW with parameters $(H,\lambda)=(0.6,\sqrt{0.09})$ withour fractional derivative (left), with fractional derivative $s=0.8$ (center) and fractional derivative $s=1$ (right). Bottom, estimated Legendre spectra, computed using  leaders (blue), $p=1$-leaders (green) and weak scaling $(\theta,\ome)$-leaders (magenta) formalims with $\theta (j) = j+j^{0.25}$ and $\omega (j)=j$ and compared to the conjectured theoretical mutlifractal spectrum (black). There is no shift in the spectra; the three spectra align with the theoretical one for all three methods (left). However, the leaders no longer align with the theoretical spectrum for a fractional derivative $s=0.6$ (center), and both the leaders and $p=1$-leaders no longer align with the theoretical spectrum for a fractional derivative $s=1$ (right).}
    \label{figmrw}
\end{figure}

Multifractal random walks are Gaussian  processes defined as  integrals of infinitely divisible stationary multifractal cascades with respect to fractional Brownian motion \cite{Bacry01,abry2009multifractal}. They have met a huge success as  models of   phenomena of multiple natures and as models on which the numerical algorithms for estimating multifractal spectra have been tested. By construction, such processes display only { \em canonical singularities} in the sense defined in \cite{Porqu2017}, i.e. their H\"older exponent, $p$-exponents and weak-scaling exponents coincide (whenever they are well defined) as a consequence of the following property:   when applying a fractional integral of order $\alpha$,  the pointwise  exponent of such processes is increased by exactly the quantity $\alpha$. This implies that the numerically  estimated spectra of the sample paths of MRWs using wavelet leaders, $p$-leaders or $(\theta, \omega)$-leaders should yield the same result, and these spectra should be shifted by $\alpha$ to the right when a fractional integral of order $\alpha$ is applied.  Figure \ref{figmrw} shows that  the  spectrum  always is correctly obtained in the case of an analysis based on the weak scaling exponent. This is in sharp contradistinction in the cases of the H\"older and the $p$-exponents  where the analysis yields a wrong spectrum when the admissibility condition for the use of the corresponding  exponent is not satisfied. 

\sloppy
\section{Multifractal analysis of brain activity measured in MEG}
\fussy

\label{secmegsig}

\subsection{Scale-free dynamics in brain activity}

Scale-free dynamics has been reported in spontaneous brain activity~\cite{He10} and in electrophysiological recordings, such as magnetoencephalography~(MEG), electroencephalography~(EEG) and local-field-potentials~(LFP)~\cite{He10,Foster16,la2018self}. The presence of scale-free dynamics in the brain was originally demonstrated in the infra-slow frequency range of the broadband spectrum~(from $0.01$~Hz to $1$~Hz~\cite{He10,Buzsaki14,He14,Becker18}) but also in the slow power fluctuations of narrow-band neuronal oscillations~\cite{Freeman00,Linkenkaer-Hansen01,Monto08,Palva13,dumeur2023multifractality}. Empirical work has revealed that scale-free dynamics of brain activity was modulated by levels of wakefulness (\textit{vs.} sleep)~\cite{Weiss09,He10,Dehghani12,Tagliazucchi13}, consciousness (\textit{vs.} anesthesia)~\cite{He09,Bartfeld15}, aging and neurodegenerative diseases~\cite{Suckling08} as well as task performance~\cite{Buiatti07,He10,He11,Ciuciu12,zilber2012modulation,zilber2013learningb,Monto08,Palva13,Lin16,la2018self}. 

The intuition behind the scale-free concept is that the relevant information in the temporal dynamics of a given signal is coded within the relations that tie together temporal scales, rather than solely in the power of neuronal oscillations in specific bands. However, its origin remains poorly understood. Brain activity recorded with MEG or EEG is more comparable to LFP, and slow dynamic fluctuations probably reflect the up and down states of cortical networks compared to spiking activity \textit{per se}~\cite{Baranauskas12}. Hence, although fast neuronal activity or avalanches can endogenously produce scale-free infra-slow brain dynamics nearby the critical regime~\cite{dumeur2023multifractality}, a careful statistical assessment remains necessary to draw conclusions on the nature of observed scale-free dynamics~\cite{Bedard06,Touboul10,Dehghani12phd}.

\subsection{Models for scale-free brain dynamics}

Scale-free dynamics recorded in electrophysiology~(MEG, EEG) has generally been quantified using a $1/f^\beta$ power spectrum model on a wide continuum of frequencies. As a result, empirical assessment has often used Fourier-based spectrum estimation. As an alternative, self-similarity provides a well-accepted model for scale-free dynamics that encompasses, formalizes, and enriches traditional Fourier $1/f^\beta$ spectrum modeling, with models such as fractional Brownian motion~(fBm) or fractional Gaussian noise (fGn)~\cite{Novikov97,He10,Ciuciu12,Ciuciu14}. The parameter of self-similarity, or Hurst exponent~$H$, matches the spectral exponent $\beta$ as $\beta = 2H-1$ for fGn and as $\beta = 2 H +1$ for fBm. In the context of brain activity, $H$ indexes how well neural activity is temporally structured (through its autocorrelation). Furthermore, although $H$ has been estimated using Detrended Fluctuation Analysis~(DFA)~\cite{Linkenkaer-Hansen01,Buiatti07,He11,Hardstone12,Palva13,Bartfeld15}, it is now well documented that wavelet-based estimators provide significant theoretical improvements and practical robustness over DFA, notably by disentangling true scale-free dynamics from non-stationary smooth trends \cite{Veitch99,Torres03,Baykut05,Ciuciu12,Ciuciu14}. For a review of statistically relevant estimations of the self-similarity parameter, interested readers are also referred to \cite{bardet2003}.

Often associated with Gaussianity, self-similarity alone does not fully account for scale-free dynamics. The main reason is that self-similarity restricts the description of neural activity to second-order statistics (autocorrelation and Fourier spectrum) and hence to additive processes. However, multiplicative processes have been proposed to provide more appropriate descriptions of neural activity \cite{Buzsaki14}. Independently of, and in addition to self-similarity, multifractality provides a framework to model these nonadditive processes \cite{Shimizu04,Suckling08,VandeVille10}. Multifractality can be conceived as the signature of multiplicative mechanisms or as the intricate combination of locally self-similar processes. For example, if a cortex patch (i.e. the anatomical resolution of MEG recordings) is composed of several small networks each characterized by a single self-similar parameter $H$, the multifractality parameter (say $M$) constitutes an index that captures the diversity of $H$s and their interactions within the patch. Qualitatively, the multifractality parameter $M$ quantifies the occurrence of transient local burstiness or non-Gaussian temporal structures, not accounted for by the autocorrelation function or by the Fourier spectrum (hence, neither by $H$ nor $\beta$). To meaningfully and reliably estimate $M$, it has been theoretically shown that the wavelet-based analysis must be extended to wavelet-leaders \cite{Wendt07} and more recently to wavelet $p$-leaders~\cite{leonarduzzi2016p}. The purpose of this section is to show that such $p$-leader formalism can fall short in certain situations in MEG time series analysis such as the presence of oscillating singularities, such as the chirps \eqref{chirp}, or when $\zeta_X(p)$ is  negative for all values of $p >0$ so that a multifractal analysis based on  $p$-exponents cannot be worked out, for any value of $p$.

\subsection{Motivations for WSE-MFA in MEG}

The development of the weak-scaling multifractal analysis is instrumental for a reliable and automated analysis of MEG times series. This statement actually results from the following key observations. 
First, from one sensor to another, MEG signals have a varying amount of regularity, some embodying
oscillating singularities. Therefore fractional integration or order $s$ has different effects on different time series.
Optimizing the order $s$ in a sensorwide manner is not tenable in practice and would mean that the input signals cannot be analyzed in a homogeneous way, or that, part of the neuronal activity is lost if we adopt the same fractional order everywhere. Second, as MEG recordings are real data, we do not have access to ground truth parameters $(H_{min}, \eta(q))$ and their estimates may be biased. 
The WSE multifractal analysis therefore allows us to avoid such inherent limitations of the standard wavelet $p$-leader formalism.

\subsection{MEG data set}

Magnetoencephalography (MEG) measures magnetic field magnitude and gradient near the surface of the skull of human subjects. The commonly received interpretation for the genesis of magnetic currents observed in MEG is that the postsynaptic currents of large neuronal assemblies of pyramidal neurons in the cortex that fire together in a synchronized manner form current dipoles whose induced magnetic field is strong enough to overcome the noise and be measured by SQUID sensors. 

We picked an ordinary resting-state recording from an openly available dataset \cite{ds004107:1.0.0} to showcase the common shortcomings of wavelet leaders and $p$-leaders in the context of state-of-the-art multifractal analysis of MEG signals. The time series were sampled at 1793 Hz, and at recording time were high-pass filtered at 0.1 Hz. We additionally low-pass filtered the data with a cutoff at 3 Hz.

\subsection{MEG signal preprocessing}

MEG signals are naturally noisy, as  sensors record every magnetic field variation, whether coming from the brain or from physiological noise sources~(e.g. eye blink, heartbeat, motion) and external ones~(e.g. power line). We followed the standard processing pipeline in order to remove the noise component in the data, making use of mne-python~\cite{Gramfort13}:

\begin{enumerate}
    \item Bad MEG sensors are identified visually.
    \item Signals coming from outside the area where the head is present are suppressed via the temporal Signal-Space Separation method (tSSS). 
    Bad channels are interpolated in the process, and head movement is cancelled by shifting to a reference position.
    \item Biological artifacts due to blinking and heartbeats are removed via Independent Component Analysis (ICA). Independent components (spatial filters) that correlate to heartbeats and blinks are identified, then the measurement is reconstructed, without the noise components.
\end{enumerate}

Further projection of the signals onto the cortical surface~(also called source localization in the field) is possible, however it is not necessary to illustrate the problems associated with low regularity in the recorded time signals: They are already present in the sensor space. 

\paragraph{Prior art.}

The low frequency fluctuations of electrophysiological time series have been shown to be approximately scale free in MEG/EEG~\cite{dehghani2010comparative,dehghani2010magnetoencephalography,zilber2012modulation}.
In particular, multifractality in MEG signals has been demonstrated to be increased in multiple brain areas during a visual discrimination task as compared to the resting state~\cite{Zilber13,la2018self} and through a multi-perceptual learning paradigm~\cite{zilber2014supramodal,zilber2014erf}. Additionally, multifractality has been observed during epileptic seizures~\cite{domingues2019multifractal} and reproduced from computational models of neural field dynamics~\cite{dumeur2023multifractality}.

\paragraph{Difficulties and aims.}
Electrophysiological recording time series are difficult to handle due to the presence of locally highly irregular singularities.
The low minimal regularity of MEG time series has required large fractional integration coefficients ($s  \geq 1.5$) to make a $p$-leaders analysis feasible with $p=2$ (the value which empirically yields the best statistical robustness (see also \cite{Wendt07}).

The lowest regularity time series is the one which sets the global integration level, as a single value of $s$ for the whole data set is required to have comparable estimates to perform statistical analysis later on.

Single outlier low-regularity time series may be ignored, annotated as bad channels and interpolated. However this carries a loss in statistical power during subsequent analyses, and should remain exceptional.

Lifting the current requirement of high fractional integration to perform multifractal analysis in neural recordings would enable a gain in sensitivity to unveil multifractality, and therefore higher statistical power in MEG data analysis.
Higher statistical power then implies being able to better determine the functional relevance of multifractality and its modulation between different experimental conditions, stimuli of patient conditions. 

\subsection{Multifractal analysis}

\label{mulanameg}

Multifractal analysis of MEG signals was performed using the open source Python Toolbox \texttt{pymultifracs}\footnote{\texttt{https://github.com/neurospin/pymultifracs}}~\cite{pymultifracs}.

\begin{figure}[htbp]
    \centering
    \includegraphics[scale=.8]{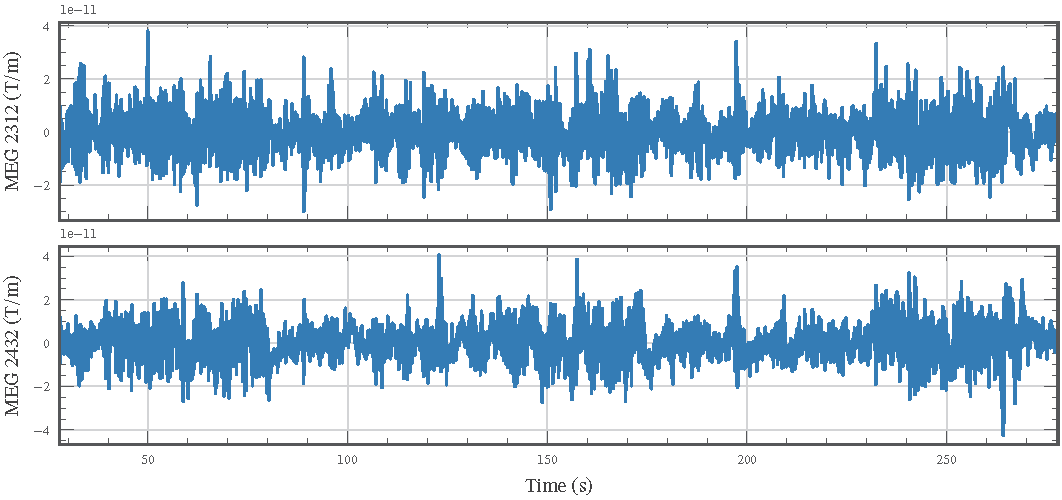}
    \caption{
    Simultaneous recording of two gradiometers: MEG~2312 and MEG~0412, single subject (magnetic field gradient in~T/m versus time in s).
    }
\end{figure}

\begin{figure}[htbp]
\centerline{  
\includegraphics[width=.75\linewidth]{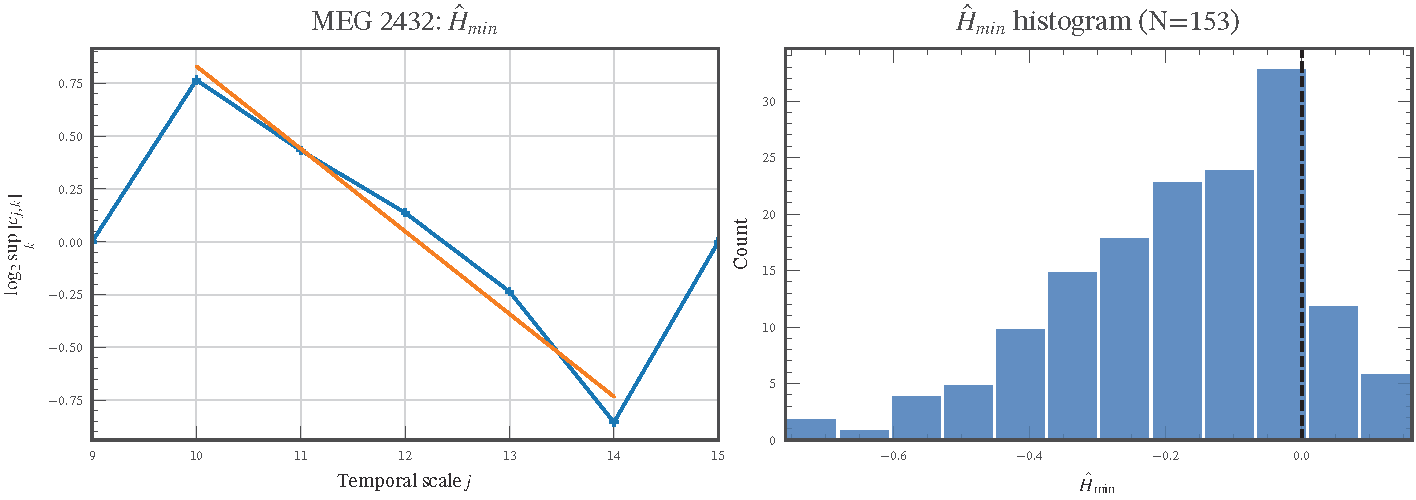}
}
    \caption{{\bf Estimation of $\Hmin$.} Left, principle of the estimation of $\Hmin$ using a log-log regression on the supremum of the wavelet coefficients, showing, for the chosen sensor, a negative estimate: 
    $\Hmin \approx -0.13 < 0$. Right: For the 153 signals with relevant multifractal behavior, we observe that for a large number of cases $\hat{H}_{min} < 0$, which precludes to perform  multifractal analysis using wavelet leaders without a priori fractional integration. \label{figRMSEHmin}}
\end{figure}

To perform multifractal analysis, the discrete wavelet coefficients $c_{j,k}$ of MEG time series are computed across scales $2^j\in [2^{j_{\rm min}}, 2^{j_{\rm max}}s]$ and time points $k/2^j$ with $j_{min},j_{max}=10, 14$ corresponding to $[10^{-2}, 1]s$, from earlier work.

\begin{figure}[htbp]
    \centerline{
    \includegraphics[width=.75\linewidth]{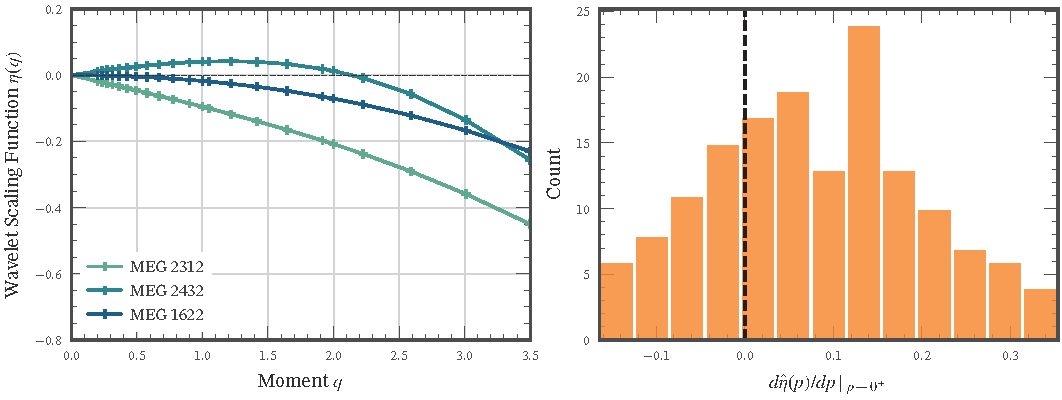}
    }
    \caption{Wavelet scaling functions for different sensors during the same recording period.  For two sensors, $\eta(q) < 0$ for all positive $q$: indicating that no $p$-leader based analysis is possible without performing a priori fractional integration.}
    \label{figStrScale}
\end{figure}

\sloppy
Among the 306 signals recorded on MEG sensors, 153 are selected as showing empirically reliable scale-free dynamics
(see \cite{leonarduzzi2014scaling} for a general methodology).

Fig.~\ref{figRMSEHmin}(left)  
illustrates the principle of the estimation $\hat{H}_{min}$ of  ${H}_{min}$ and reports Fig.~\ref{figRMSEHmin}(right) the empirical distribution of estimated $\hat{H}_{min}$. 
It shows that for the selected MEG signals, a large proportion of $\hat{H}_{min}$ are estimated negative, which implies that the wavelet leader formalism is not applicable.
\sloppypar

Fig.~\ref{figStrScale})(left) shows the functions $\eta(p)$ estimated from different sensors: 
When there exists a $p>0$ such that $\eta(p) > 0$, the $p$-leader based multifractal formalism can be applied without fractional integration.
Conversely, when $\forall p>0, \eta(p) < 0$ (equivalently, $\frac{d\hat{\eta}(p)}{dp}\big|_{p\to 0^+}<0$) the $p$-leader based multifractal formalism cannot be applied without prior fractional integration.
Fig.~\ref{figStrScale}(right) reports the histogram of  $\frac{d\hat{\eta}(p)}{dp}\big\vert_{p\to 0^+}$  and shows that for a significant subset of sensors, $\frac{d\hat{\eta}(p)}{dp}\big\vert_{p\to 0^+}<0$ and thus that there is no value of $p$ that allows the $p$-leader formalism to be used without fractional integration.

These two observations motivate the use of the WSE-formalism to perform mutlifractal analysis with no recourse to fractional integration. 

\begin{figure}[htbp]
    \centering
    \includegraphics[width=\linewidth]{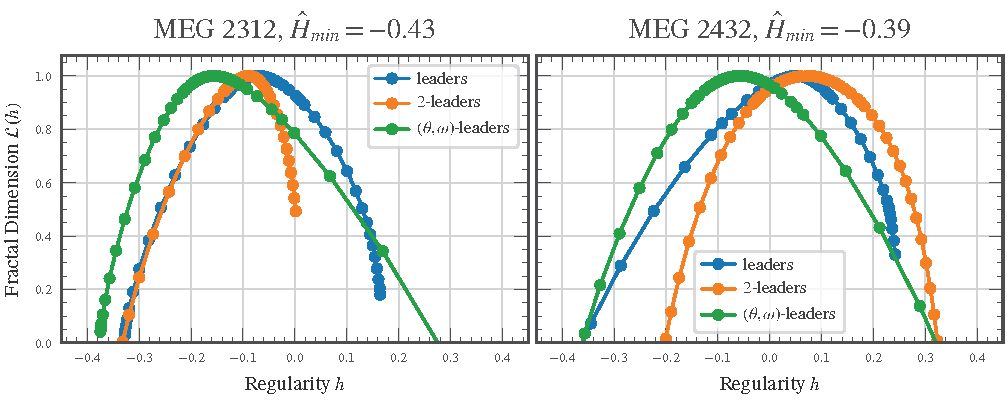}
    \caption{Legendre spectra, estimated from three different regularity exponent and multiscale quantities: H\"older exponent and leaders (blue),  $p=2$-exponent and  $p=2$-leaders (orange), and WS-exponent and WS-leaders (green, using $\theta(j) = j + j^{0.4}$), for two different sensors (right and left).
    Integration with $s=1$ is used for the leaders, and for the $2$-leaders on the left plot.
    When integration is used, the spectra are offset by the integration factor: $h \to h-1$.
    The WSE spectrum reveals lower regularity in both cases thus indicating more accurate multifractal analysis.}
    \label{figSpect1}
\end{figure}

As an illustration, multifractal spectra for sensors MEG~2312 and MEG~0412 are compared in Fig.~\ref{figSpect1}, for formalisms based on different multiscale quantities: Leaders, $p=2$-leaders, and WSE. 
Time series were fractionally integrated ($s=1$) for leaders and $p=2$-leaders, whenever required (i.e., when $\hat{H}_{min} < 0$ for leaders and $\eta(2) < 0$ pour $2$-leaders).
Fig.~\ref{figSpect1}(left) corresponds to the case where the use of either leader and 2-leader requires fractional integration ($s=1$), whereas the WSE-based spectrum can be estimated without fractional integration.
Fig.~\ref{figSpect1}(right) corresponds to the case where the use of leader requires fractional integration ($s=1$), whereas the $2$-leaders and  WSE-based spectra can be estimated without fractional integration.
In both cases, spectra are ploted with the shift $h \rightarrow h-s $ that \emph{cancels} the simple translation effect $h \rightarrow h+s $ induced by fractional integration, that would correspond to the ideal case where fractional integration does not alter in other ways the estimation of the multifractal properties of data. 
Both cases illustrate that the WSE spectrum reveals lower regularity in data than leaders and $2$-leaders permit to do, thus illustrating that fractional itegration may alter or impair an accurate estimation of the multifractal properties in real-world data. 

\begin{figure}[htbp]
    \centering
    \includegraphics[width=.75\linewidth]{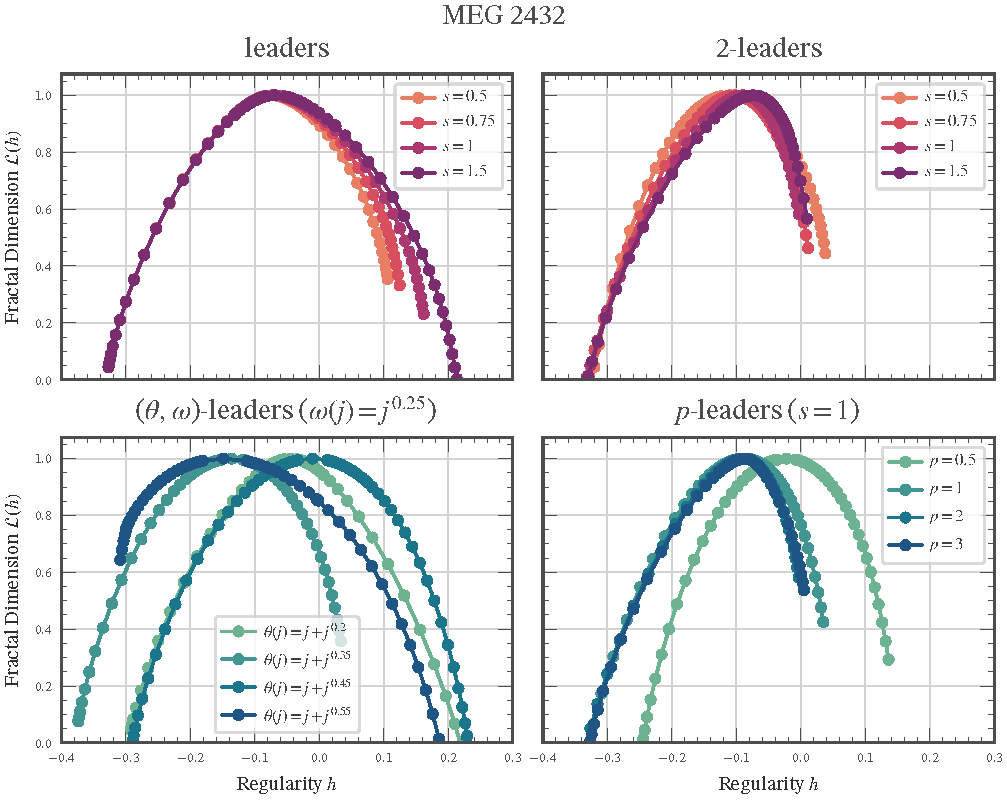}
    \caption{{\bf Legendre spectra for sensor MEG2432.}
    Leader-based (top left) and $p=2$-leader-based estimates for different fractional integration orders $s$.
   Bottom left,
   WS-leader-based estimates for different $\theta(j) = j + j ^\beta $.
   Bottom right, 
   $p$-leader-based estimates for different $p$ and with and fractional integration of order $s=1$.
   When integration is used, the spectra are offset by the integration factor: $h \to h-s$.}
    \label{fig:parameters}
\end{figure}

Finally, Fig.~\ref{fig:parameters} shows the impact of the a priori user-chosen parameters on estimated spectra for the time series collected on one same sensor (MEG 0412).
Top row plots illustrate the sensitivity of using increasing fractional integration order $s$ on both leaders and $2$-leaders spectra. 
Bottom row(right) plot illustrates the sensitivity of varying $p$ in $p$-leaders spectra.
Bottom row(left) plot illustrates the sensitivity of varying $\theta = j + j^\beta$ in the $(\theta, \omega)$-leaders.
Under mild hypotheses, this choice should theoretically be without impact on estimation asymptotically.

To summarize, we have analyzed 306 signals of which only 153 showcase multifractality with reliable scaling dynamics. However, out of the 153 signals, 130 have an estimated $\hat{H}_{min}<0$, within which there are 40 signals for which analysis using $2$-leaders is not feasible. The analysis by WSE now makes it possible to overcome this obstacle.

\section{Conclusion}

Estimating $H_{min}$ via linear regression is difficult in the context of limited or noisy data, and may lead to incorrect guesses about the degree of fractional integration required to obtain sensible Legendre spectra. Furthermore, in some experimental cases, $\eta(p)$ varies on a signal-by-signal basis, which in the absence of WSE-based analysis would require either different fractional integration coefficients, or more realistically to suffer from over-integrating part of the time series.

The WSE formalism mitigates these difficulties in dealing with time series of varying $H_{min}$, by providing a homogeneous method to deal with time series that have heterogeneous multifractal properties.

 \clearpage
 \bibliographystyle{plain}
 \bibliography{biblio.bib}

\end{document}


\begin{figure}[ht]
    \begin{minipage}{.49\linewidth}
        \includegraphics[scale=1, left]{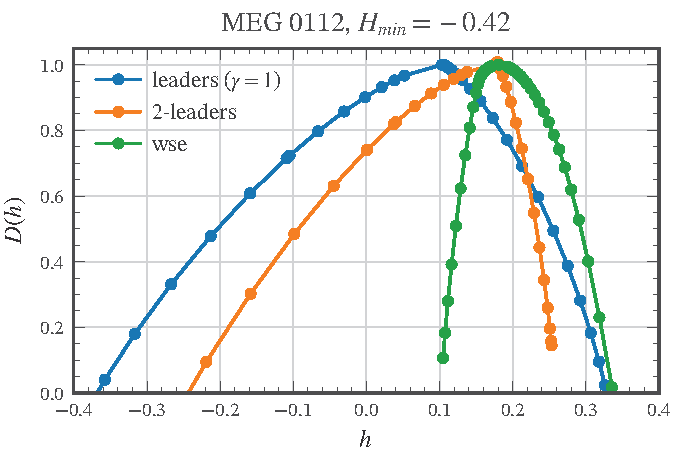}
        \includegraphics[scale=1, left]{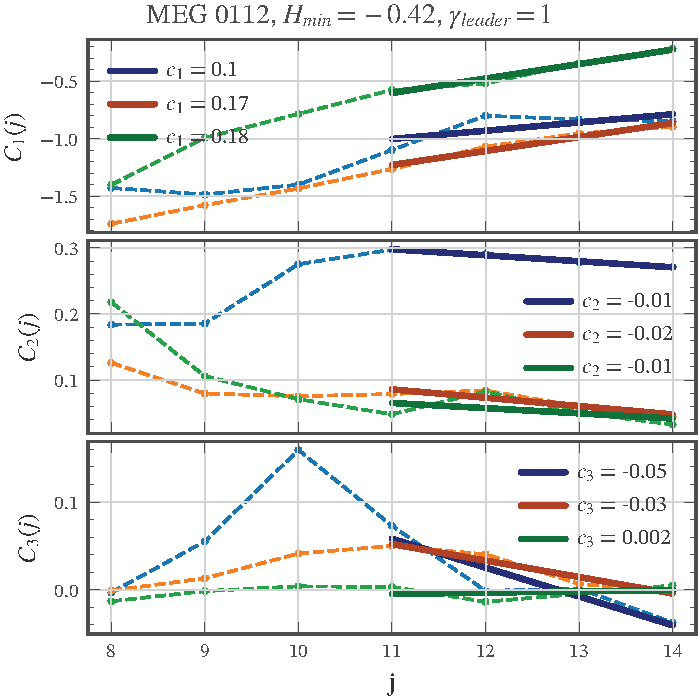}
    \end{minipage}%
    \begin{minipage}{.49\linewidth}
        \includegraphics[scale=1, right]{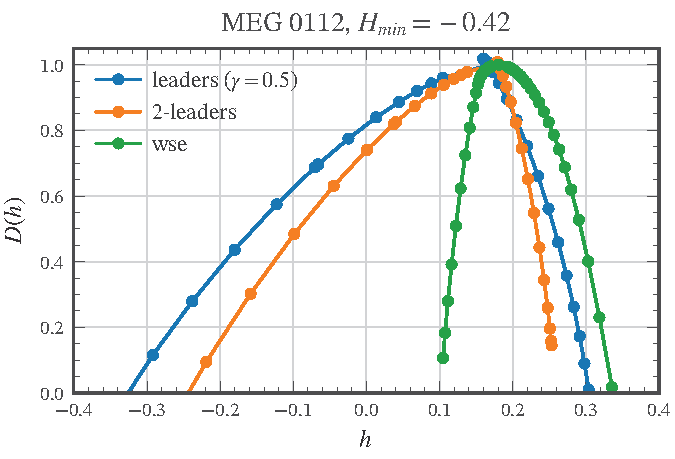}
        \includegraphics[scale=1, right]{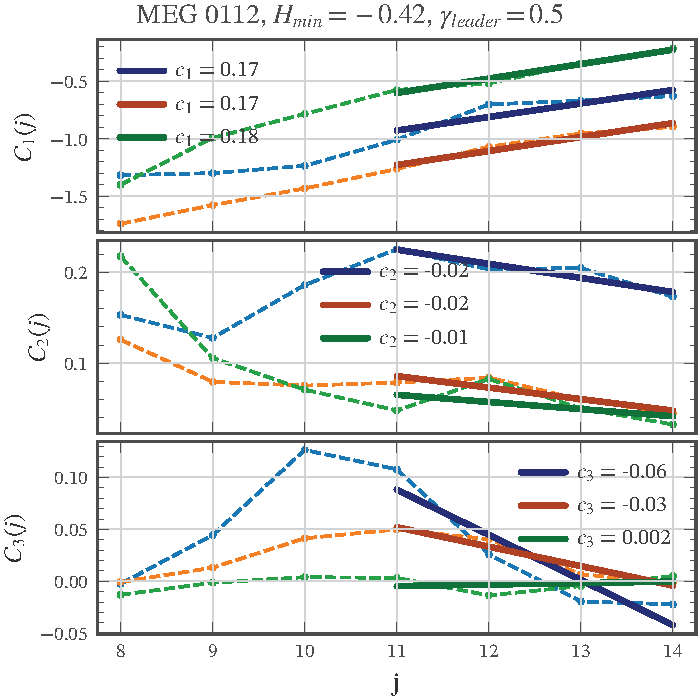}
    \end{minipage}%
    \label{figSpect1}
\end{figure}

\clearpage
\newpage

\begin{figure}[ht]
    \centering

    \begin{minipage}{.49\linewidth}
        \raggedleft
        \includegraphics[scale=1, left]{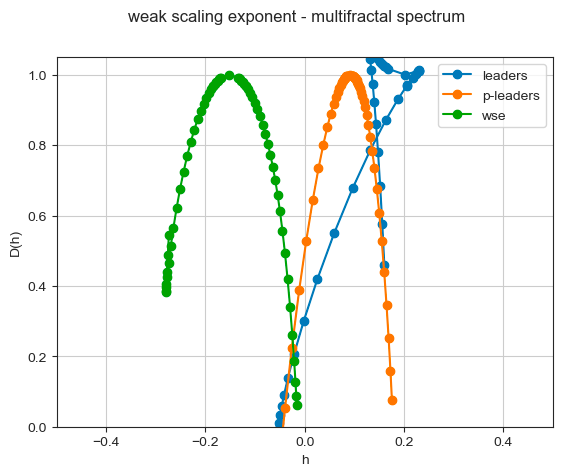}
        \includegraphics[scale=1, left]{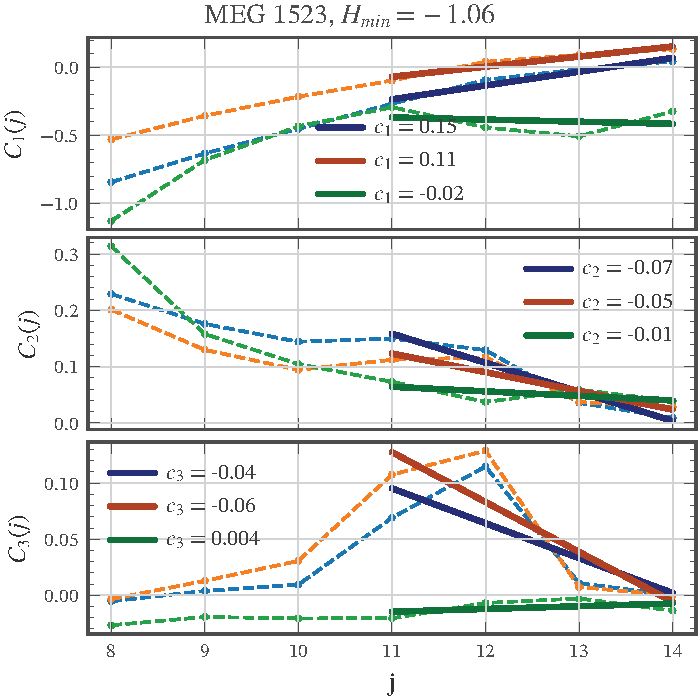}
    \end{minipage}%
    \begin{minipage}{.49\linewidth}
        \raggedright
        \includegraphics[scale=1, right]{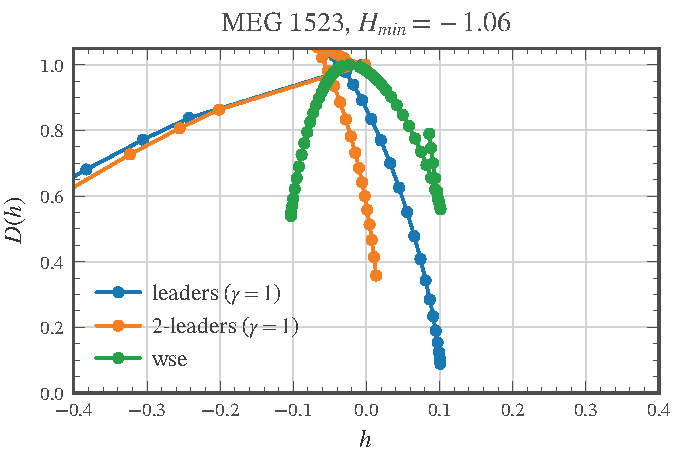}
        \includegraphics[scale=1, right]{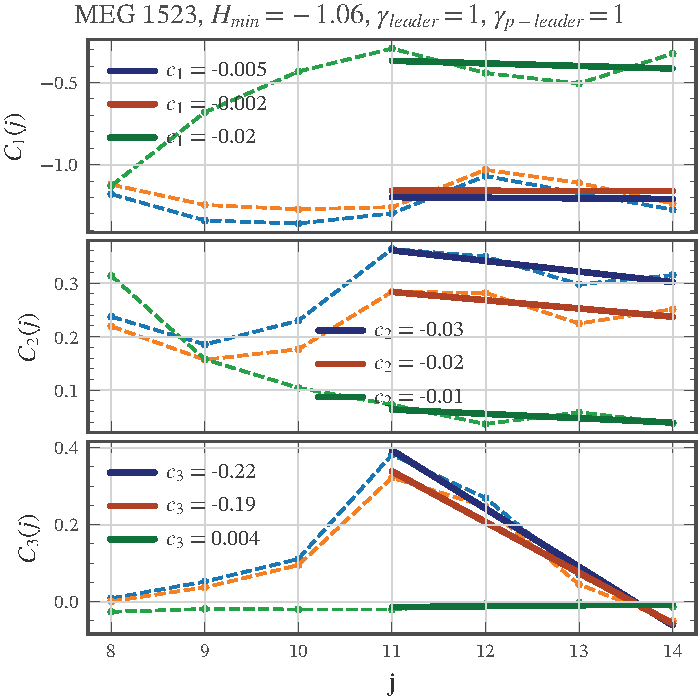}
    \end{minipage}%
    
    \label{figSpect1}
\end{figure}

\clearpage
\newpage

\begin{figure}[ht]
    \centering

    \begin{minipage}{.49\linewidth}
        \raggedleft
        \includegraphics[scale=1, left]{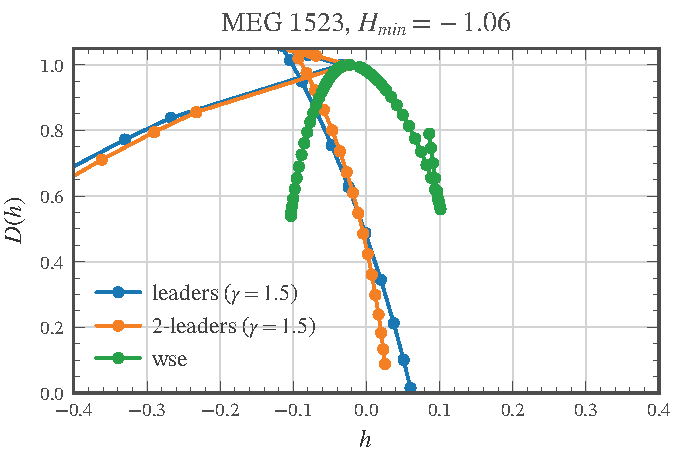}
        \includegraphics[scale=1, left]{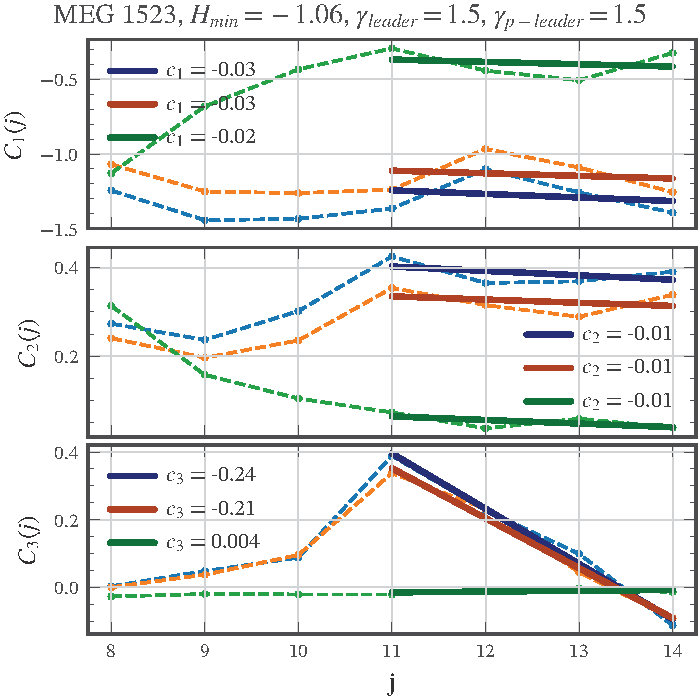}
    \end{minipage}%
    \begin{minipage}{.49\linewidth}
        \raggedright
        \includegraphics[scale=1, right]{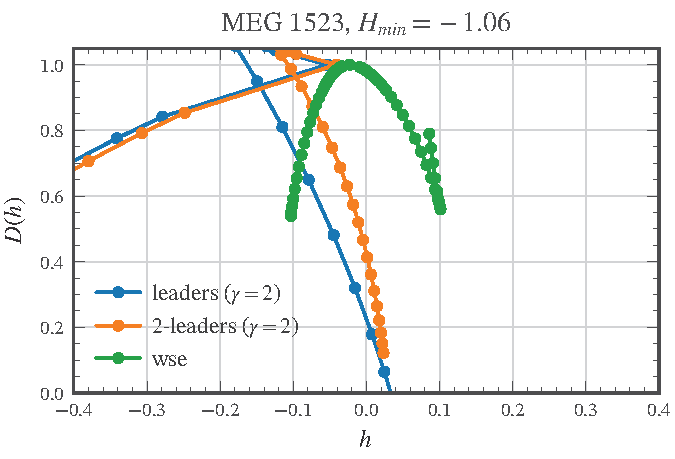}
        \includegraphics[scale=1, right]{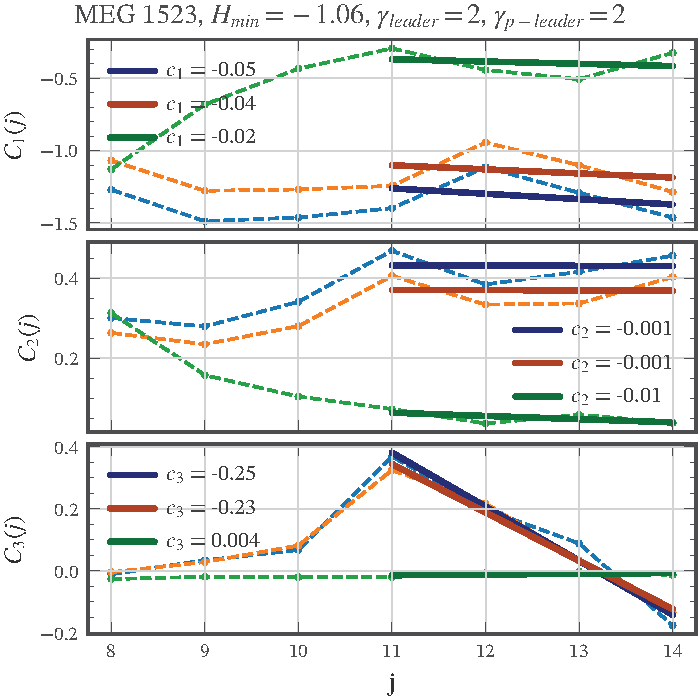}
    \end{minipage}%
    
    \label{figSpect1}
\end{figure}